\documentclass[notitlepage,aps,prb,reprint,twocolumn, longbibliography,superscriptaddress]{revtex4-1}
\usepackage{graphicx}
\usepackage{float} 
\usepackage{wrapfig}
\usepackage{amsmath}
\usepackage{amssymb}
\usepackage{braket} 
\usepackage{comment} 
\usepackage[colorlinks, allcolors=blue]{hyperref}
\usepackage[all]{hypcap}
\usepackage[mathlines]{lineno}
\usepackage{ulem}
\usepackage{color}

\begin{document}

\title{Bloch Oscillations Along a Synthetic Dimension of Atomic Trap States}
\author{Christopher Oliver}
\affiliation{School of Physics and Astronomy, University of Birmingham, Edgbaston, Birmingham B15 2TT, United Kingdom}
\author{Aaron Smith}
\affiliation{School of Physics and Astronomy, University of Birmingham, Edgbaston, Birmingham B15 2TT, United Kingdom}
\author{Thomas Easton}
\affiliation{School of Physics and Astronomy, University of Birmingham, Edgbaston, Birmingham B15 2TT, United Kingdom}
\author{Grazia Salerno}
\affiliation{Department of Applied Physics, School of Science, Aalto University, Espoo, Finland}
\author{Vera Guarrera}
\affiliation{School of Physics and Astronomy, University of Birmingham, Edgbaston, Birmingham B15 2TT, United Kingdom}
\author{Nathan Goldman}
\affiliation{Center for Nonlinear Phenomena and Complex Systems, Université Libre de Bruxelles (U.L.B.), B-1050 Brussels, Belgium}
\author{Giovanni Barontini}
\affiliation{School of Physics and Astronomy, University of Birmingham, Edgbaston, Birmingham B15 2TT, United Kingdom}
\author{Hannah M. Price}
\affiliation{School of Physics and Astronomy, University of Birmingham, Edgbaston, Birmingham B15 2TT, United Kingdom}

\begin{abstract}
    Synthetic dimensions provide a powerful approach for simulating condensed matter physics in cold atoms and photonics, whereby a set of discrete degrees of freedom are coupled together and re-interpreted as lattice sites along an artificial spatial dimension. However, atomic experimental realisations have been limited so far by the number of artificial lattice sites that can be feasibly coupled along the synthetic dimension. Here, we experimentally realise for the first time a very long and controllable synthetic dimension of atomic harmonic trap states. To create this, we couple trap states by dynamically modulating the trapping potential of the atomic cloud with patterned light. By controlling the detuning between the frequency of the driving potential and the trapping frequency, we implement a controllable force in the synthetic dimension. This induces Bloch oscillations in which atoms move periodically up and down tens of atomic trap states. We experimentally observe the key characteristics of this behaviour in the real space dynamics of the cloud, and verify our observations with numerical simulations and semiclassical theory. The Bloch oscillations thus act as a smoking gun signature of the synthetic dimension, and allow us to characterise the effective band structure. Our methods provide an efficient approach for the manipulation and control of highly-excited trap states, and set the stage for the future exploration of topological physics in higher dimensions through the use of a tunable artificial gauge field and finite-range interactions.
\end{abstract}

\maketitle

\section{Introduction}
Synthetic dimensions provide a powerful approach for simulating condensed matter physics in cold atoms~\cite{Boada2012, Celi2014, Mancini2015, Stuhl2015, Livi2016, Kolkowitz2017, Chalopin2020,lienhard2020realization,Kanungo2021, Gadway2016, Viebahn2019, cai2019experimental,Sundar2018, kang2020creutz} and photonics~\cite{lustig2019photonic,dutt2020single,balvcytis2021synthetic,chen2021real}, and they are opening up many new avenues for simulating and exploring exotic physics, including  quantum Hall ladders~\cite{Mancini2015, Stuhl2015,Gadway2016}, non-Hermitian topological bands~\cite{Wang2021}, topological Anderson insulators~\cite{meier2018observation}, and even lattice physics in four dimensions or higher~\cite{Viebahn2019, Boada2012,price2015}. A key reason that this framework is so powerful is that it is very general, and can be applied to a wide-range of very different physical systems. For example, synthetic dimensions have so far been realised experimentally in cold atoms with hyperfine~\cite{Boada2012, Mancini2015, Stuhl2015}, magnetic~\cite{Chalopin2020}, Rydberg~\cite{lienhard2020realization,Kanungo2021}, and clock states~\cite{Livi2016, Kolkowitz2017}, as well as with momentum~\cite{Gadway2016,  Viebahn2019}, orbital~\cite{kang2020creutz} and superradiant states~\cite{cai2019experimental}. However, in these experiments, the size of the synthetic dimension has been so far limited by the number of states that can feasibly be coupled. Indeed, the largest momentum state lattice~\cite{An2021} employed so far consists of a 1D lattice of 21 sites.

Notably, it has been recently realised that such limitations can be lifted if external degrees of freedom associated with trapping potentials are used to generate the synthetic dimensions~\cite{Price2017, Salerno2019, lustig2019photonic}. Indeed, trapping potentials typically allow for tens or hundreds of trapped states in each direction, and by suitably coupling them one could implement very long synthetic dimensions, unleashing the full potential of this technique for quantum simulation. Additionally, this kind of synthetic dimension is extremely appealing because it provides a framework for the manipulation and control of trap states. A range of applications including quantum simulations in optical lattices~\cite{Wirth2011, Muller2007, Li2016}, trapped and guided atom interferometry~\cite{Hu2018,Frank2014,Guarrera2015} and quantum thermodynamics~\cite{Vin2016,Quan2007,Uzdin2015} require the use of highly-excited trapped states, which are generally difficult to realise with a good degree of precision and control.

In this work, we experimentally engineer a very long synthetic dimension of many tens of atomic trap states by dynamically modulating the harmonic trap of an ultracold atomic sample~\cite{Price2017, Salerno2019}. 
By controlling the driving frequency we generate a force along the synthetic dimension that induces Bloch oscillations, which act as a smoking gun signature that the synthetic dimension behaves as expected. Bloch oscillations were first famously predicted for electrons moving in a crystal under an electric field, and have since been observed in various setups, including optical lattices for cold atoms~\cite{Dahan1996}, as well as synthetic dimensions of photonic frequency modes~\cite{bersch2009experimental} and of room-temperate molecular angular momentum states~\cite{Floss2015}. However, Bloch oscillations in our experiment are physically very different from previous realizations, as they correspond to atoms periodically oscillating between low- and high-energy states of the harmonic trap. As such, another benefit of the synthetic dimension Bloch oscillations implemented here is that they allow us to explore highly-excited harmonic states, and thus can lead towards a novel approach for quantum engineering of external atomic states. More generally, this work paves the way for the exploration of higher-dimensional quantum Hall physics with artificial magnetic fields, and opens new opportunities in quantum simulations more widely. For example, a tunable artificial gauge field can be implemented by spatial modulation of the phase of the driving potential, allowing access to 2D quantum Hall physics. Furthermore, mean-field interactions between the atoms in real space should give rise to exotic interactions that are long-range and decay with distance along the synthetic dimension~\cite{Price2017}, in contrast to the usual interactions in atomic gases and to the $SU(N)$ interactions in some other atomic synthetic dimension schemes. We therefore expect interesting ground state physics under the inclusion of interactions which will be of interest for future study.
\section{Overview of the scheme}
\begin{figure*}
	\includegraphics[width=0.99\textwidth]{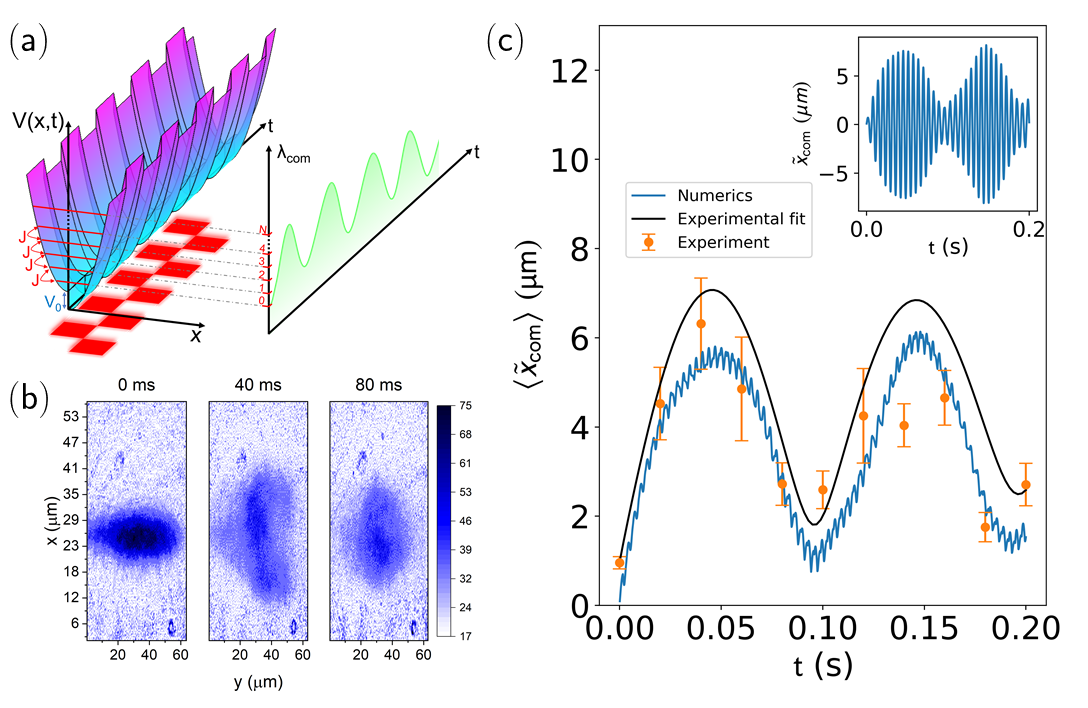}
	\centering
	\caption{{\bf Schematic of the experiment and signatures of the synthetic-dimension Bloch oscillations in real space.} (a): Schematic of the DMD pattern (\textit{red}), which shakes the harmonic trap (\textit{blue/purple}) and couples together nearest-neighbour trap states with an approximately uniform hopping amplitude, $J$, in order to create the synthetic dimension. Bloch oscillations (\textit{green}) can be driven along the synthetic dimension by applying a force along $\lambda$, corresponding to detuning the shaking frequency from the trap frequency. (b): Experimentally, the cloud is imaged after a short time-of-flight expansion, as demonstrated here for a detuning of $\Delta = 9.84 \times 2\pi$ Hz. The colour scale is the column density in arbitrary units. The full dynamics includes micromotion within each driving period, so the experimental data are averaged over several values of $\varphi$. This averaging procedure makes the cloud appear to widen, although the true cloud width is approximately constant. (c): The real-space center-of-mass position, $x_{com}$, of the atomic cloud is extracted, as shown here for the same data as in (b). The experimental data ({\it orange}) are fitted ({\it black}) with a function motivated by Bloch oscillations in the synthetic dimension (see Appendices B and D), which captures the real-space dynamics well. Also shown are numerical results ({\it blue}) with a suitable TOF correction, discussed in the Main Text. The micromotion within each driving period can be seen in the inset for an initial driving phase of $\varphi=0$. To reduce these unwanted micromotion effects, we average both the experiment and numerics over several values of $\varphi$ (e.g. $\varphi = 0$ and $\pi/2$) to obtain the results shown in the main panel. As can be seen numerically, the residual micromotion oscillations have a small amplitude, which can be further reduced by averaging over more initial driving phases. Panels (b) and (c) use experimental parameters of $V_0 = 4.16 \text{ nK}$, $T = 20 \text{ nK}$, $\omega_y = 10 \times 2\pi \text{ Hz}$, and $\omega_x = 166.5 \times 2\pi \text{ Hz}$, where the latter is determined experimentally by shifting the oscillation frequency data to pass through $(|\Delta|, f) = (0, 0)$. Experimental errorbars are 1$\sigma$ statistical errors.}
	\label{fig:F1}
\end{figure*}

To introduce our scheme, let us consider an atomic cloud in a cigar-shaped harmonic trap with trap frequencies $\omega_x = \omega_z \gg \omega_y$. In order to realise the synthetic dimension, we couple together the atomic trap states of the strong trapping potential along $x$ with the spatially- and temporally-varying driving potential given by:
\begin{equation}
    V_D(x, t) = - V_0 \Theta(x \sin (\omega_D t + \varphi))  \label{eqn:drive}  
\end{equation} 
where $V_0$ is the driving amplitude, $\Theta(x)$ is the Heaviside step function, $\omega_D$ is the driving frequency and $\varphi$ is the initial driving phase. Physically, this corresponds to illuminating the upper half of the atomic cloud with constant power for the first half of the period, $T_D= 2\pi / \omega_D$, before illuminating the lower half with the same constant power over the second half of the period (see Fig.~\ref{fig:F1}(a)). This driving protocol is chosen because it is simple to implement, robust and it leads to a simple Floquet Hamiltonian with near-constant nearest-neighbour hoppings, corresponding to a textbook 1D tight binding model, as discussed below.
Combining Eq.~\ref{eqn:drive} with the 1D harmonic oscillator Hamiltonian along $x$ gives the time-dependent Hamiltonian:
\begin{equation}
    \hat{H}(t) = \hbar\omega_x \sum_{\lambda}\lambda\ket{\lambda}\bra{\lambda} + V_D(x, t),
    \label{eqn:H_t}
\end{equation}
written in the eigenstate basis of the ``strong'' trap along $x$, as indexed by $\lambda = 0, 1, 2 ...$. 
\begin{figure*}
	\includegraphics[width=0.99\textwidth]{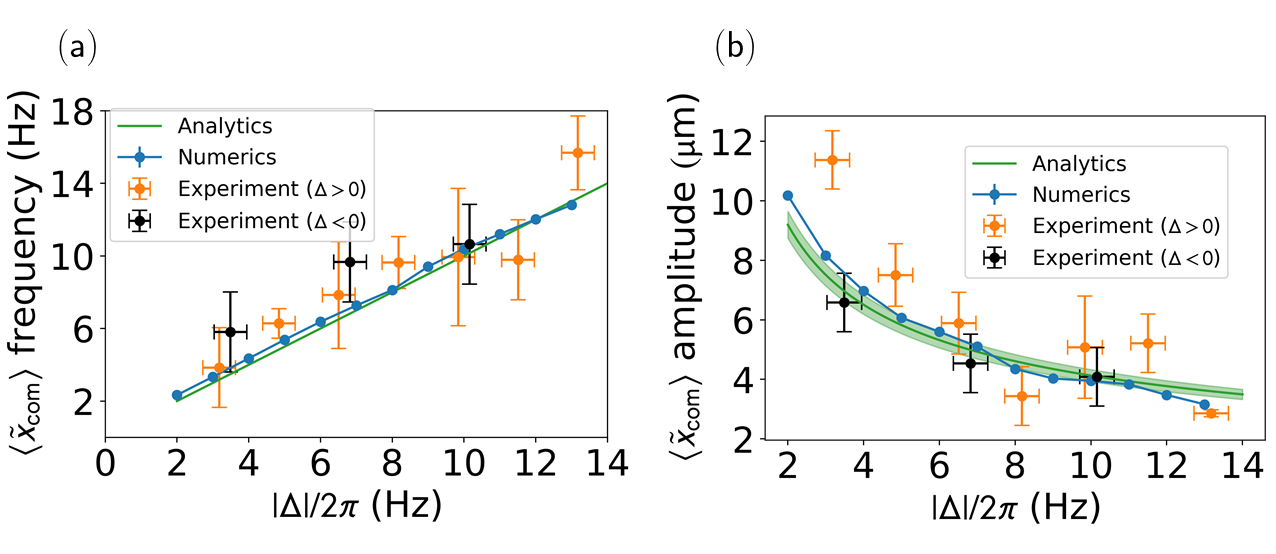}
	\centering
	\caption{{\bf Bloch oscillation frequency and real space amplitude as a function of detuning} (a): Frequency of the Bloch oscillations for the experiment ({\it orange} for $\Delta>0$ and {\it black} for $\Delta <0$) and for numerics ({\it blue}). The observed trend is in agreement with the analytical prediction ({\it green}) of $f_B= |\Delta| / 2 \pi $ for Bloch oscillations. (b): Amplitude in real space of Bloch oscillations for the same data as plotted in panel (a). The analytical prediction ({\it green}) shows the expected real-space amplitude as calculated under appropriate approximations from the synthetic-dimension Bloch oscillations (see Appendix G), with the green error band calculated from the errors on $J$ and other numerical parameters. The numerical results are obtained by fitting the same function as in experiment to the TOF-corrected, RMS-averaged results like in Fig.~\ref{fig:F1}(c). Note that the numerical fit parameters also include error bars, but these are smaller than the datapoint size. We use the same parameters as Fig.~\ref{fig:F1}. Experimental errorbars are 1$\sigma$ statistical errors.}
	\label{fig:F1_part2}
\end{figure*}
\begin{figure*}
	\includegraphics[width=0.99\textwidth]{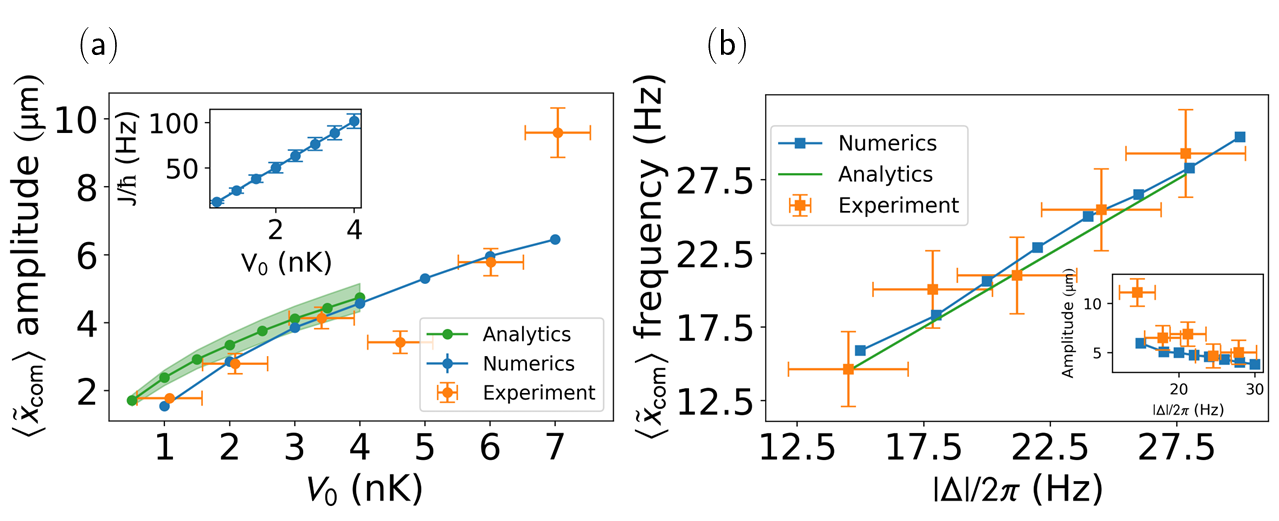}
	\centering
	\caption{{\bf Varying the shaking power to control the dynamics.} (a): The real-space amplitude [c.f Fig.~\ref{fig:F1_part2}(b)] for different shaking powers at a fixed detuning of $\Delta = 8.3 \times 2\pi \text{ Hz}$, with $\omega_x = 162.6 \times 2\pi \text{ Hz}$. As can be seen, the experiment ({\it orange}), numerics ({\it blue}) and analytics ({\it green}) increase with the shaking power; this is because, as shown in the inset, the nearest-neighbour hopping amplitude, $J$, from Floquet theory (see Appendix F), increases with $V_0$ and hence the Bloch-oscillation amplitude along the synthetic dimension increases. Analytics are only shown up to $V_0 = 4$ nK, as at higher shaking potentials the simple nearest-neighbour tight-binding model in Eq.~\eqref{eqn:H} is no longer a good description (see Appendix F), although the numerics and experiment still appear to exhibit Bloch-oscillation dynamics. This is shown further in (b), where we plot the amplitude fit parameter (inset) and the frequency fit parameter, for $V_0 = 11.96 \text{ nK}$, with $\omega_x = 142.1 \times 2\pi \text{ Hz}$, $\Delta < 0$, and other parameters as in Fig.~\ref{fig:F1} and~\ref{fig:F1_part2}. Despite the large shaking power, we still observe similar trends to that at low power [c.f. Fig~\ref{fig:F1_part2}], and still with agreement in panel (b) to the $f_B= |\Delta| / 2 \pi $ analytical result from the simple tight-binding model. Experimental errorbars are 1$\sigma$ statistical errors. Note that all numerical datapoints include error bars, but these are smaller than the datapoints. The green error band on the analytics is calculated from the errors on $J$ and other numerical parameters.}
    \label{fig:F2}
\end{figure*}
The stroboscopic dynamics of this system is captured by an effective time-independent Floquet Hamiltonian, which we can 
approximate over a large number of trap-states by (see Appendix F):
\begin{equation}
    \hat{\mathcal{H}} \approx \hbar\Delta \sum_{\lambda} \lambda \ket{\lambda}\bra{\lambda}  + J \sum_{\lambda} \left[ ie^{ i \varphi} \ket{\lambda + 1}\bra{\lambda} + \text{h.c} \right] \label{eqn:H}
\end{equation} 
where $\Delta \equiv \omega_x - \omega_D$ is the (small) drive detuning and $J$ is a uniform hopping amplitude, which itself depends on $V_0$ and $\omega_x$ and is calculated using Floquet theory (Appendix F). Note that this description is valid for near-resonant driving in a deep harmonic trap, i.e. such that $\omega_x \simeq \omega_D \gg \Delta, J/\hbar$.  As depicted in Fig.~\ref{fig:F1}(a), Eq.~\ref{eqn:H} describes a particle hopping between nearest-neighbour sites along a 1D tight-binding lattice with unit spacing in a synthetic dimension. The detuning plays the role of a constant force, $F\equiv - \hbar \Delta$, which therefore can induce Bloch oscillations. Note that the shaking phase $\varphi$ appears in the effective Hamiltonian as a hopping phase, which we will exploit below to average over unwanted micromotion effects.

In the absence of a force along the synthetic dimension, i.e. when $\Delta=0$, the effective model in Eq.~\ref{eqn:H} is translationally invariant along the synthetic dimension and has a single energy band in the Brillouin zone. Applying a nonzero force (i.e. $\Delta \neq 0$) accelerates a semiclassical wavepacket formed in the synthetic dimension bulk, such that it undergoes Bloch oscillations across the Brillouin zone, with a center-of-mass (COM) position along the synthetic dimension, ${\lambda}_{\text{com}}$, that varies as~\cite{Price2017}:
\begin{eqnarray}
    \lambda_{\text{com}} (t) = \lambda^0_{\text{com}} + \frac{2J}{\hbar\Delta}(1 - \cos(2\pi f_Bt)), \label{eqn:lambda}
\end{eqnarray}
from the initial position $\lambda_{\text{com}}(t\! =\! 0) \!= \!\lambda^0_{\text{com}}$ and where $f_B$ is the Bloch oscillation frequency. As we can set the spacing between the fictitious synthetic lattice sites equal to one, the periodic Brillouin zone covers $[-\pi, \pi ]$. The Bloch oscillation frequency is then set by the magnitude of the applied force divided by the length of this Brillouin zone, i.e. $f_B= |\Delta| / 2 \pi $, and so is entirely controlled by the detuning. Conversely, the amplitude of the Bloch oscillations is proportional to the bandwidth divided by the force, i.e. $4 J / \hbar \Delta$, and so depends on the detuning but also, through $J$, on the trap frequency and shaking power (Appendix F). The Bloch oscillations therefore provide a way to transport atoms between different trap states, with independent control over both the timescale and number of trap states explored.
\begin{figure}
	\includegraphics[width=0.45\textwidth]{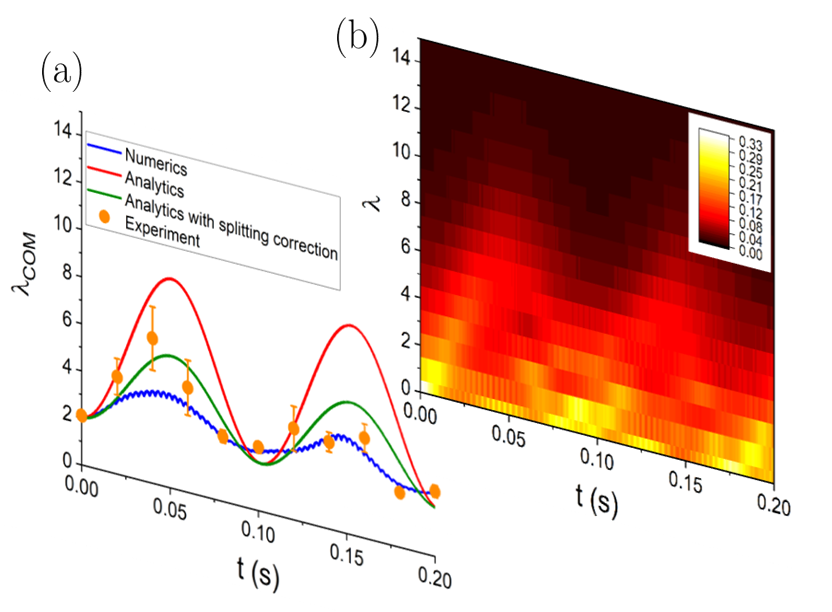}
	\centering
	\caption{{\bf Example Bloch oscillation in synthetic space.} (a): Evolution of the COM, $\lambda_{\text{com}}$, and the numerical density distribution heatmap along the synthetic dimension (b) for the same data as Fig.~\ref{fig:F1}(b), for $\varphi = 0$. Both Bloch oscillations and high-frequency micromotion are visible in the 2D numerics ({\it blue line}) for a non-interacting thermal cloud (Appendix C). We also convert the corresponding real space experimental data to synthetic space under suitable assumptions (see Appendix E). The observed Bloch oscillation frequency is in good agreement with the 1D analytical semiclassics (Eq.~\ref{eqn:lambda}, {\it red line}), albeit with a lower amplitude as only around half the atoms from the thermal cloud oscillate along the synthetic dimension [c.f. density distribution heatmap in (b)]. To correct for this, the analytical result is rescaled ({\it green line}) as discussed in the text. The heatmap shows the numerical atomic density (integrated along the $y$ direction) with respect to $\lambda$ over the oscillation, showing the cloud-splitting. Both panels use experimental parameters of $\omega_x = 166.5 \times 2\pi \text{ Hz}$, $\Delta = 2\pi \times 9.84$ Hz, $V_0 = 4.16 \text{ nK}$, $T = 20 \text{ nK}$, and $\omega_y = 10 \times 2\pi \text{ Hz}$. Experimental errorbars are 1$\sigma$ statistical errors.}
	\label{fig:F3}
\end{figure}
Experimentally, we use a thermal cloud of $^{87}$Rb atoms in a harmonic trap with trapping frequencies $\omega_x = \omega_z \simeq 2\pi \times 160$ Hz and $\omega_y \simeq  2\pi\times 10$ Hz. The cloud was measured, both in-situ and with standard time-of-flight techniques, to have an initial temperature of $T \simeq 20\text{ nK}$. We used a rapid evaporation ramp to prevent the sample from condensing at this temperature, thus reducing the effect of mean-field interactions~\cite{Price2017}, which may complicate the dynamics and will be of interest in future investigations. To realise the driving potential of Eq.~\ref{eqn:drive}, we utilise a digital micromirror device (DMD) that allows us to dynamically and spatially control the intensity profile of a laser beam with wavelength 800 nm (see Appendix A). We verify that the trapping and driving potentials are aligned to within $\simeq 1\mu$m. Our driving potential was chosen because it was found to be the most effective and robust, in the sense of not being sensitive to misalignments and other imperfections. This is in addition to the favourable theoretical properties discussed above. We then perform absorption imaging of the atomic cloud in position space after a very short time-of-flight (TOF) expansion of $t_{\text{tof}} = 5\text{ ms}$, chosen to increase the visibility of the dynamics. This is demonstrated in Fig.~\ref{fig:F1}(b) for a detuning of 9.84 Hz, where we plot the real space cloud density as a function of time for an example Bloch oscillation, showing the cloud COM being displaced away from the origin. Note that the density is averaged over several values of $\varphi$ in order to reduce the effect of micromotion. As such, the cloud appears to widen along $x$ in time, although the cloud width is actually approximately constant. The initial temperature of $T \simeq 20\text{ nK}$, corresponds to $\lambda_{\text{com}}^0\approx 2$ for an initial Maxwell-Boltzmann distribution of atomic energies. This means that, initially, the atoms start near one ``edge" of the synthetic dimension (at $\lambda = 0$), and so atoms must move up the synthetic dimension, irrespective of the sign of the detuning, i.e. the direction of the force. The cloud also does not have a Gaussian distribution with respect to the synthetic dimension as the above semiclassical theory implicitly assumes; nevertheless, as we shall show, the prediction in Eq.~\ref{eqn:lambda} works well once appropriate corrections are included (see Appendix G).
\section{Experimental and Theoretical Results}
Bloch oscillations in $\lambda$-space naturally translate into atomic motion along $x$ in real space, as different harmonic oscillator eigenstates have different real-space profiles. The resulting motion can be seen in Fig.~\ref{fig:F1}, where we report the measured dynamics of the real-space COM position under the action of the shaking potential, as a function of time. It is important to notice that the full dynamics also includes micromotion within each driving period~\cite{Price2017}; for the real-space COM, as shown numerically in the inset of Fig.~\ref{fig:F1}(c), the micromotion corresponds to large and fast oscillations as the atoms slosh backwards and forwards in the trap, while the stroboscopic Bloch oscillations translate into variations in the envelope of the dynamics. {We are not able to reliably achieve the high time resolution to observe the micromotion in experiment, so we apply an averaging procedure to remove it. This is because of drifts in experimental parameters on a timescale of hours,  which would make the large number of measurements required to reconstruct fast dynamics impractical. To overcome this, we root-mean-square (RMS) average over different experimental runs with suitably-chosen different initial driving phases, $\varphi$. The raw cloud images in Fig.~\ref{fig:F1}(b) have themselves been averaged in this way. This has the effect of making the cloud appear to widen significantly over the oscillation, although in each single shot the width is approximately constant. Note that this averaging procedure slightly lowers the apparent amplitude of the motion. However, by reducing the micromotion effects, we can then clearly observe the real-space signatures of the synthetic-dimension Bloch oscillations, as reported in Fig.~\ref{fig:F1}(c) for the same parameters as Fig.~\ref{fig:F1}(b) (\textit{orange}), where the experimental data is fitted by a function (\textit{black}) motivated by synthetic-dimension Bloch oscillations (see Appendices B and D):
\begin{equation}
    x(t) = A\sqrt{1 - e^{-gt}\cos(2\pi ft + \phi)},
    \label{eqn:fit}
\end{equation}
where we fit to find the amplitude $A$ and frequency $f$. We introduce the additional fit parameters $g$ and $\phi$ to capture some details of the experimental data (see Appendices B and D). As can be seen, this fit captures the behaviour of the data very well, with agreement between the experiment and numerical simulations ({\it blue curve}) of a non-interacting 2D thermal cloud (see Appendix C). {Our numerical simulations use the time-dependent Hamiltonian to evolve an ensemble of states, each of which is a superposition over the eigenstates of the 2D trap with random phase factors to destroy the phase coherence. Physical observables such as the cloud density are found by averaging over these phases. To account for the fact that we do not measure the true position of the cloud due to the TOF expansion, we include an approximate TOF correction to the cloud centre-of-mass position in our numerical simulations. In particular, we use the simulated COM momentum  $p_{\text{com}}$ to find the semiclassical cloud velocity, and then shift the simulated cloud COM at each timestep (Appendix D).

To further characterise our experimental results, we plot in Fig.~\ref{fig:F1_part2}(a) the values of the oscillation frequency obtained by fitting our data for different detunings. As can be seen, for both experiment and numerics, the frequency increases linearly with detuning, as expected from the analytical Bloch-oscillation frequency ({\it green line}) given by $f_B = |\Delta| / 2\pi$ in both real and synthetic space [c.f. Eq.~\ref{eqn:lambda}]. The trapping frequency is determined by shifting a linear fit to the measured oscillation frequencies to pass through $(|\Delta|, f) = (0, 0)$. This provides a straightforward way to measure the trapping frequency, but does mean that any systematic uncertainty would not be detected. 

In Fig.~\ref{fig:F1_part2}(b) we show how the amplitude of the real-space motion depends on the detuning by plotting the amplitude fitting parameters. As can be seen, the experiment ({\it orange/black}) and numerics ({\it blue}) both clearly show the expected growth in the real-space amplitude as the detuning decreases and higher-excited atomic trap states are explored. To make a quantitative comparison with semiclassical Bloch oscillations (Eq.~\ref{eqn:lambda}), we have derived an analytical expression (see 
Appendix D) that converts the expected Bloch oscillation amplitude from synthetic space to real space under appropriate assumptions, including a correction for the finite fraction of atoms participating in the dynamics, as discussed further below. The expression is based on the formula:
\begin{equation}
   x_{\text{com}} = \sqrt{\lambda_{\text{com}} -\sigma_x^2 + \frac{1}{2}},
    \label{eqn:conversion_formula}
\end{equation}
which connects the real space cloud COM $x_{\text{com}}$ and width $\sigma_x$ to the synthetic space COM $\lambda_{\text{com}}$ under certain assumptions. This result is derived in Appendix E. This analytical prediction is
plotted in Fig.~\ref{fig:F1_part2}(b) ({\it green}), with errors calculated from our numerical parameters. As can be seen, there is agreement between the experiment, numerics and the analytics, demonstrating that we have achieved good control of the synthetic-dimension Bloch oscillations.

We can also independently increase the number of atomic trap states explored (i.e. the Bloch oscillation amplitude) while keeping the oscillation frequency constant, by increasing the shaking power, $V_0$, and hence the hopping parameter $J$ [c.f. Eq.~\ref{eqn:H}]. The dependence of the real-space COM amplitude on $V_0$ is shown in Fig.~\ref{fig:F2}(a) for a fixed detuning $\Delta = 8.3 \times 2\pi \text{ Hz}$, while the inset shows the variation of the hopping $J$, as calculated with Floquet theory (Appendix F). As can be seen, the amplitude in experiment (\textit{orange}), numerics (\textit{blue}) and analytics (\textit{green}) all increase as the hopping increases [c.f. Eq.~\ref{eqn:lambda}]. Note that the analytical result is only plotted for $V_0 \leq 4$ nK, as at higher shaking powers, our simple analytical model (Eq.~\ref{eqn:H}) breaks down (see Appendix F). 
Despite this, we still observe clear Bloch oscillation dynamics at high power. This is further demonstrated in Fig.~\ref{fig:F2}(b), where we use a very strong potential of $V_0 = 11.96 \text{ nK}$, and still observe the same qualitative trends (i.e. the amplitude decreasing with the detuning and the frequency being equal to the detuning) as in the low power regime.
\begin{figure}
	\includegraphics[width=0.45\textwidth]{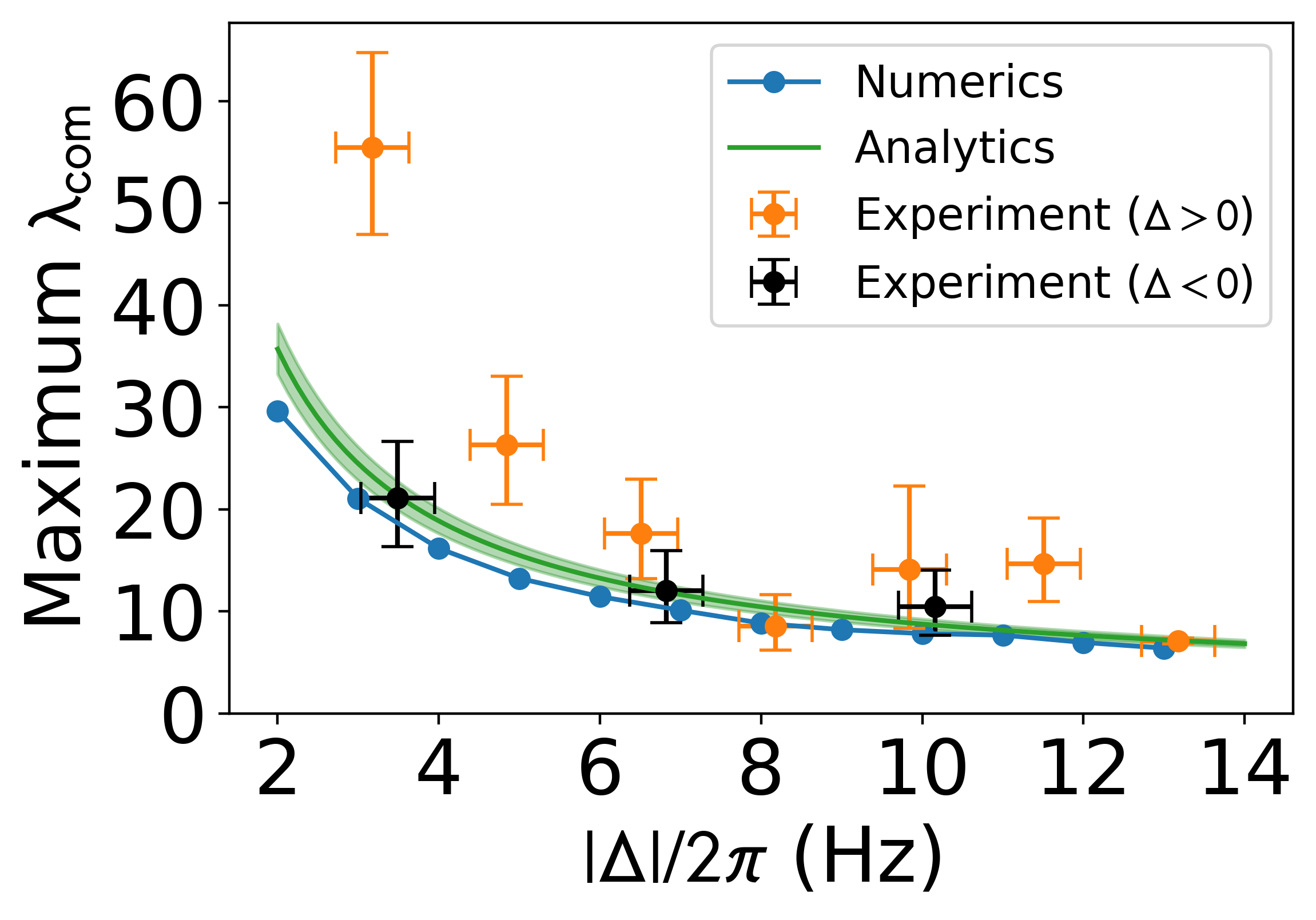}
	\centering
	\caption{{\bf Bloch oscillation amplitude in synthetic space.} The extrapolated maximum value of $\lambda$ reached by the oscillating part of the cloud as a function of $\Delta$, demonstrating that we have a long, controllable synthetic dimension. As discussed in Appendix H, we convert the experimental fit parameters in Fig.~\ref{fig:F1_part2}(b) to synthetic space (\textit{orange/black}), correcting for cloud splitting. We also plot our numerical maximum $\lambda$ values (\textit{blue}) for $\varphi = 0$, again corrected for cloud splitting. Furthermore, we derive appropriate analytics (\textit{green}) for comparison (see Appendix G). Error bars on experimental points were calculated by converting the real space errors to synthetic space, while the error bars on the analytics were found by propagating errors on $J$ and other numerical parameters. We use experimental parameters of $\omega_x = 166.5 \times 2\pi \text{ Hz}$, $V_0 = 4.16 \text{ nK}$, $T = 20 \text{ nK}$, and $\omega_y = 10 \times 2\pi \text{ Hz}$.}
	\label{fig:F4}
\end{figure}
Finally, we can visualise the Bloch oscillations along the synthetic dimension more directly, by translating our real-space experimental measurements [from Fig.~\ref{fig:F1}(c)] into $\lambda$-space, under suitable approximations (see Appendix G). These experimental results are plotted in Fig.~\ref{fig:F3}(a) ({\it orange}), along with numerical simulations ({\it blue curve}). Note that the latter is not averaged over different initial driving phases; however, in synthetic space, the micromotion is only a fraction of a ``lattice spacing" and becomes negligible as the trapping frequency $\omega_x$ increases~\cite{Price2017}.

We also compare our experimental results directly with the semiclassical analytical predictions [Eq.~\ref{eqn:lambda}] with ({\it green curve}) and without ({\it red curve}) multiplying by a constant numerical re-scaling factor to account for the fraction of atoms contributing to the dynamics (Appendices G and H). This correction was also used in Fig.~\ref{fig:F1_part2} and~\ref{fig:F2} and is necessary because the COM is skewed downwards by the thermal cloud splitting into two distinct parts, with approximately half of the atoms remaining in low $\lambda$-states during the oscillation, as can be seen in the numerical density distribution heatmap in Fig.~\ref{fig:F3}(b). This effect is likely caused by small oscillations in the Floquet Hamiltonian matrix elements, and we discuss methods to reduce this splitting effect in Appendix I}.

Importantly, we can also convert our experimental amplitude fit parameters (Fig.~\ref{fig:F1_part2}(b)) to synthetic space and divide by the constant numerical cloud splitting factor in order to extrapolate how far up $\lambda$ the oscillating part of the cloud is actually exploring (see Appendix H). This is plotted in Fig.~\ref{fig:F4} (\textit{orange/black}), where we can see that the fraction of the cloud involved in the Bloch oscillations is moving up several tens of Bloch states, demonstrating that we are creating a very long synthetic dimension. This is supported by our numerical (\textit{blue}) and analytical results (\textit{green}) (see Appendices C and D). Note that the number of sites along the synthetic dimension will eventually be limited experimentally by anharmonicities in the trapping potential~\cite{Price2017}; however, in this experiment, this does not play an important role. We included the appropriate quartic anharmonic terms in our numerical simulations and found that they had no effect in the range of trap states that are relevant here. We discuss the properties of the excited states we create, and methods to improve their fidelity, in Appendix I. Finally, we note that our methods can be extended towards single-site resolution to on the order of or below $\hbar\omega_x/k_B$. That would start the dynamics with only the $\lambda = 0$ state populated. The subsequent dynamics would maintain the single $\lambda$-state character. This point is discussed further in Apprendix I.
\section{Conclusions}
Our experimental results show that we have engineered a synthetic dimension of atomic trap states by projecting a time-dependent shaking potential onto an atomic cloud via a digital micro-mirror device. Through control of the shaking potential's detuning, we induced Bloch oscillations along the synthetic dimension, observing the key characteristics of these dynamics in the real-space motion of the cloud. Our experiment demonstrates that a long and controllable synthetic dimension can be created. This opens up the way towards the exploration of topological physics using a synthetic dimension of harmonic trap states~\cite{Price2017,Salerno2019} by introducing a controllable artificial gauge field using a spatially-varying shaking phase. The spatio-temporal control of the shaking potential can also allow for future investigations of phenomena such as magnetic barriers, as well as the controlled population of excited atomic trap states, including direct imaging of the states~\cite{Scherer2010} and single-site resolution of our current methods (Appendix I). Moreover, the addition of mean-field interactions in the cloud should lead to exotic interactions along the synthetic dimension~\cite{Price2017} and, in turn, interesting ground state physics.

\section*{Appendix A: Experiment}

In our experimental sequence we load $^{87}$Rb atoms from a 3d MOT into a crossed optical dipole trap and then perform forced evaporative cooling \cite{DanielNJP}. The final trapping frequencies are $f_x=f_z\simeq160$ Hz, and $f_y\simeq10$ Hz, where $z$ is the vertical axis, resulting in a cloud elongated along $y$, with $N\simeq2\times10^4$ atoms at $\simeq$20 nK. We conclude a posteriori that the degeneracy of the $x$ and $z$ trapping directions does not affect the dynamics because of the good agreement between theory and experimental data. We also verified using horizontal imaging that there are no significant dynamics along the $z$-direction. The optical setup to realise the dynamical potential and high resolution imaging is described in detail in Ref.~\onlinecite{Smith2021}.
In brief, the light produced by a 800 nm laser is reflected by a DMD and then sent onto the atoms along the vertical direction, using an optical setup that produces a demagnification of a factor 100. The DMD is an 2d array of $1920\times1080$ micromirrors, each with size 10.8 $\mu$m. Each micromirror can be individually tilted every 100 $\mu$s, allowing us to produce dynamical optical potentials. The atoms are imaged on a CCD mounted in the vertical direction using a 20$\times$ magnifying system with a resolution of $\simeq2$ $\mu$m. The numerical aperture of our imaging system is 0.28. We verify the alignment between the driving and trapping potentials to a precision of $\simeq$1 $\mu$m by imaging both the cloud and the DMD pattern at once. This was done every 20 runs to avoid slow drifts. We also note that we do not observe significant heating of the cloud due to the driving potential. We determine this by observing the width of the cloud in the weak trapping direction after a short TOF expansion and finding that it does not change, as shown by experimental data in Fig.~\ref{fig:density}. This is caused by any ``heated'' atom spilling out of the weak trap, leading to some atom loss but a fixed temperature.

\section*{Appendix B: Real space experimental data analysis}

To reduce micromotion effects, the experimental data for the real-space COM position is RMS-averaged over initial driving phases drawn randomly from $2\pi n/30$, with $n = 0, 1, ... 30$. (The effects of RMS-averaging are discussed further in Appendix D) The resulting data are then fitted to the function: 
\begin{equation}
    x(t) = A\sqrt{1 - e^{-gt}\cos(2\pi ft + \phi)} ,
    \label{eqn:fitting}
\end{equation}
where $A$ is the amplitude, $f$ is the frequency, $g$ is a damping factor and $\phi$ is a phase offset which accounts for random variation in the state preparation. The functional form of this fitting function is motivated by translating the semiclassical prediction for synthetic-dimension Bloch oscillations (Eq.~\ref{eqn:lambda}) into real space. As shown in Appendix E, under certain approximations, this conversion can be achieved by taking:
 \begin{equation}
   x_{\text{com}} = \sqrt{\lambda_{\text{com}} -\sigma_x^2 + \frac{1}{2}},
\end{equation}
 where $ x_{\text{com}}$ and $\sigma_x$ are, respectively, the COM and width of the cloud (in harmonic oscillator lengths) with respect to the real position coordinate, $x$. Note that in choosing the form of the fitting function, we assume that the cloud width $\sigma_x$ is approximately constant as a function of time, as has also been verified numerically (Appendix D). The fitting parameters $f$ and $A$ are then plotted, for example, in Fig.~\ref{fig:F1_part2}(a) and (b) respectively. 

\section*{Appendix C: Details of Numerical Simulations}
\begin{figure}
    \centering
    \includegraphics[width=0.5\textwidth]{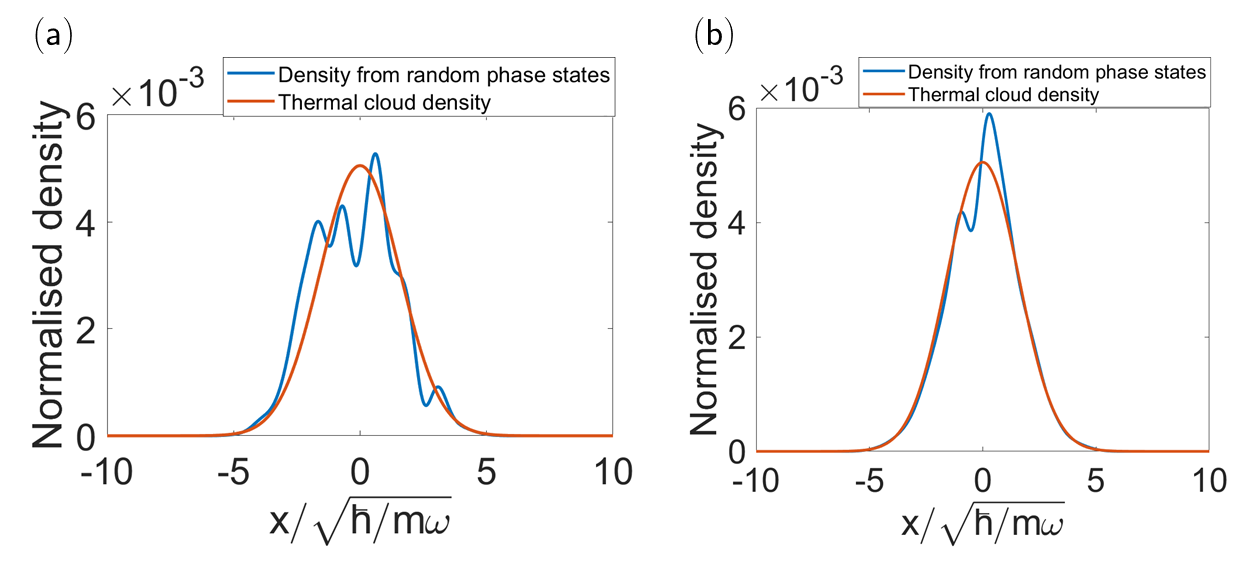}
    \caption{The 1D atomic density obtained from random-phase-state averaging (\textit{blue}) compared to the known density of a thermal cloud (\textit{red}) (Eq.~\ref{eq:thermal}) with temperature $T = 20 \text{nK}$ in a harmonic trap of frequency $\omega_x = 2\pi \times 166.5 \text{ Hz}$, and $N=16$. In (a), we average over 5 random phase states, and in (b) over 100 states. As can be seen, (a) shows significant fluctuations which decrease with averaging over more states, as shown in (b).}
    \label{fig:rp}
\end{figure}
As shown in the main text, we numerically simulate the motion of a thermal cloud in two dimensions under the time-dependent Hamiltonian $\hat{H}(t)$ (Eq. 2 in the main text, with the $y$-terms restored). In so doing, we choose to work in the $\lambda - y$ basis; this avoids discretising the $x$-direction and hence reduces the size of the matrix representing the Hamiltonian in the numerics. From our numerical simulation of the wave-function, we then calculate the cloud density $\rho(\lambda, y, t)$, and convert this into real space to yield $\rho(x, y, t)$. The time evolution of the wave-function is done by numerically time-evolving an appropriate initial state as:
\begin{align}
    \ket{\psi(t + dt)} = \exp{\left(-\frac{i\hat{H}(t)dt}{\hbar}\right)}\ket{\psi(t)},
\end{align}
where $dt$ is a sufficiently small timestep. In order to simulate the dynamics of a non-interacting thermal cloud (i.e. a non-interacting gas which is distributed over the levels of the trap according to a classical Boltzmann distribution), we choose an initial state of the form~\cite{Ponomarev2008, Gelman2006}:
\begin{equation}
    \ket{\psi_\mathbf{\theta}} = A\sum_{i=0}^{N-1} \sqrt{p_i}\exp{(i\theta_i)}\ket{\phi_i},
\end{equation}
where A is a normalisation factor, $p_i = \exp{(-\beta E_i)} / Z$ is the Boltzmann weight for the $i^{\textrm{th}}$ eigenstate of the 2D harmonic trap $\ket{\phi_i}$ with energy $E_i$, $\beta$ is the inverse temperature, $Z$ is the partition function for the 2D harmonic trap and $\theta_i$ is a random phase drawn from a uniform distribution between 0 and $2\pi$. Note that a finite number, $N$, of harmonic trap states is used in this construction; in order to safely neglect higher-energy trap states, we check numerically that the Boltzmann weight has decayed to a sufficiently small value.

In the numerics, we then sequentially generate $N'$ such random phase states, each with a different set of random phases. Each state is then time-evolved and the resultant densities are averaged together. By averaging over random phase factors, we destroy the phase coherence of the state; omitting this step would correspond to selecting a particular (and likely unphysical) coherent superposition of trap states.
\begin{figure*}[t!]
    \centering
    \includegraphics[width=\textwidth]{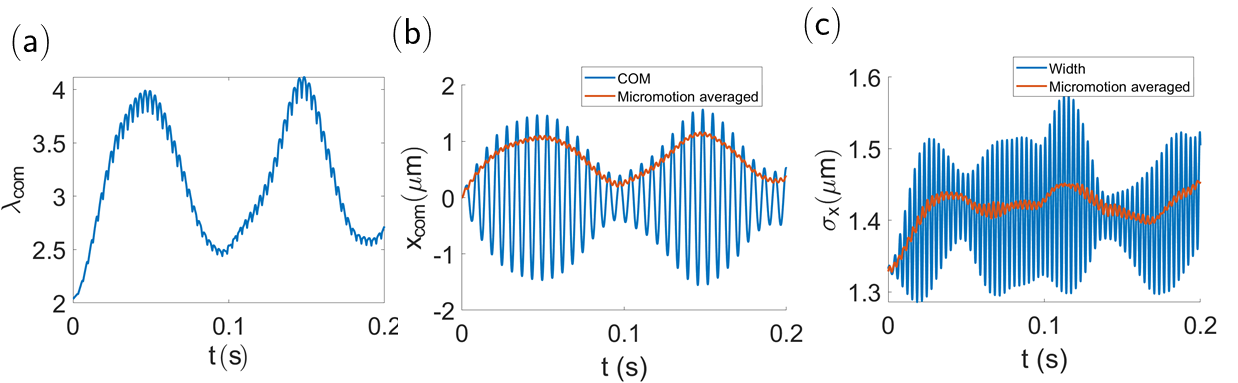}
    \caption{Example unprocessed data from the numerical simulations corresponding to Fig. 1(b) and 3(a) and (b) in the main text. (a): synthetic dimension COM as a function of time for $\varphi = 0$, showing the Bloch oscillation and micromotion. (b): real space COM, also showing micromotion under the Bloch oscillation envelope (\textit{blue}), together with the root-mean-square average over a pair of initial driving phases $0$ and $\pi / 2$ (\textit{red}). (c): real space cloud width (\textit{blue}) and its micromotion average (\textit{red}), which is approximately constant in time to one decimal place. The phase-averages are used in Section V to verify the result derived there. We use the parameters: $V_0 = 4.16 \text{ nK}, \omega_x = 166.5 \times 2\pi \text{ Hz}, \Delta = 9.84 \times 2\pi \text{ Hz}, \omega_y = 10 \times 2\pi \text{ Hz}, T = 20 \text{ nK}$, as in Fig. 1, 2, 3 and 4 in the main text.}
    \label{fig:da_unprocessed}
\end{figure*}
To further illustrate the importance of the random-phase averaging, we show how this can reproduce the expected density for a thermal cloud at $t = 0$. The latter can be found from the system's density matrix $ \hat{\rho} = \sum_ip_i\ket{\phi_i}\bra{\phi_i}, 
$
as:
\begin{align}
    \begin{split}
        \rho(\mathbf{r}) &= \bra{\mathbf{r}}\hat{\rho}\ket{\mathbf{r}} = \sum_ip_i|\phi_i(\mathbf{r})|^2,
    \end{split}
\end{align}
where $\phi_i(\mathbf{r}) = \braket{\mathbf{r}|\phi_i}$~\cite{Pethwick2001}. We can also calculate the density of the random phase state as:
\begin{align}
    \begin{split}
        \rho_{\mathbf{\theta}}(\mathbf{r}) &= \braket{\psi_\mathbf{\theta}|\mathbf{r}}\braket{\mathbf{r}|\psi_\mathbf{\theta}} \\
        &= \sum_{ij}\sqrt{p_ip_j}\exp{(i(\theta_i - \theta_j))\braket{\phi_j|\mathbf{r}}\braket{\mathbf{r}|\phi_i}}.
    \end{split}
\end{align}
which, as can be seen, involves a double sum over the harmonic trap states. However, averaging over the random phases gives:
\begin{equation}
    \rho(\mathbf{r}) = \frac{1}{(2\pi)^{N'}}\int_{0}^{2\pi}\prod_{i = 1}^{N'}d\theta_i\rho_{\mathbf{\theta}}(\mathbf{r})  = \sum_ip_i|\phi(\mathbf{r})|^2,
\end{equation}
as desired, where we used the identity:
\begin{equation}
    \int_{0}^{2\pi}\exp{(i(\theta_i - \theta_j))}d\theta_i = 2\pi\delta_{ij}.
\end{equation}
The random phase state therefore reproduces the density for a thermal cloud at $t = 0$ under suitable averaging. 

We can also demonstrate the effects of phase-averaging numerically as shown in Figure~\ref{fig:rp}, where this technique is applied to a 1D harmonic trap for (a) only five random phase states and (b) for 100 random phase states. In both cases, the blue curve is the density calculated via the above method and the orange curve is the expected thermal cloud density calculated explicitly as:
\begin{equation}
    \rho(x) = Ae^{-\frac{x^2}{2\sigma^2}}, \label{eq:thermal}
\end{equation}
where $A$ is a normalisation constant and $\sigma = k_BT/m\omega_x^2$ is the cloud width, controlled by the trap frequency and temperature~\cite{Pethwick2001}. For large enough numbers of random phases states included in the average (Fig.~\ref{fig:rp}(b)), we see good agreement with the expected thermal cloud density. Note that in so doing, we have assumed that our cloud is non-interacting, because if interactions are present, we can no longer use a Boltzmann-weighted superposition over single-particle trap states. Throughout this work, we use 50 random phase states in our numerics.

\section*{Appendix D: Details of Numerical Data Analysis}

Here we describe the data analysis steps carried out on the simulated cloud density (Appendix C) in order to extract the Bloch oscillation frequency and amplitude, which we have then compared to experiment and analytical results in the main text. 

Firstly, the cloud center-of-mass (COM) and width are calculated from the real space cloud density $\rho(x, y, t)$, found using the method in Appendix C. An example is shown in Fig.~\ref{fig:da_unprocessed}, where we see oscillations in both the synthetic and real space COM ((a) and (b) respectively), including high-frequency micromotion, as discussed in the main text. Note that the $\lambda$-space COM is calculated as $\lambda_{\text{com}} = \sum_{\lambda}\lambda\rho(\lambda, t)$, where $\rho(\lambda, t)$ is the probability density with respect to the synthetic dimension, calculated numerically as in Appendix C, with the $y$-dependence integrated out. We also see that, although the cloud does visit higher harmonic oscillator states, the cloud width (panel (c)) is approximately constant in time. We use this observation in Appendix G to simplify our analysis. Throughout this section, we use the typical experimental parameters: $V_0 = 4.16 \text{ nK}, \omega_x = 166.5 \times 2\pi \text{ Hz}, \Delta = 9.84 \times 2\pi \text{ Hz}, \omega_y = 10 \times 2\pi \text{ Hz}, T = 20 \text{ nK}$ and the initial driving phase $\varphi = 0$ for any data including micromotion and $\varphi = 0$ and $\pi/2$ for any data where micromotion has been averaged out, as discussed further below. These are the same parameters as in Fig. 1, 2, 3 and 4 in the main text.
\begin{figure}[t!]
	\includegraphics[width=0.45\textwidth]{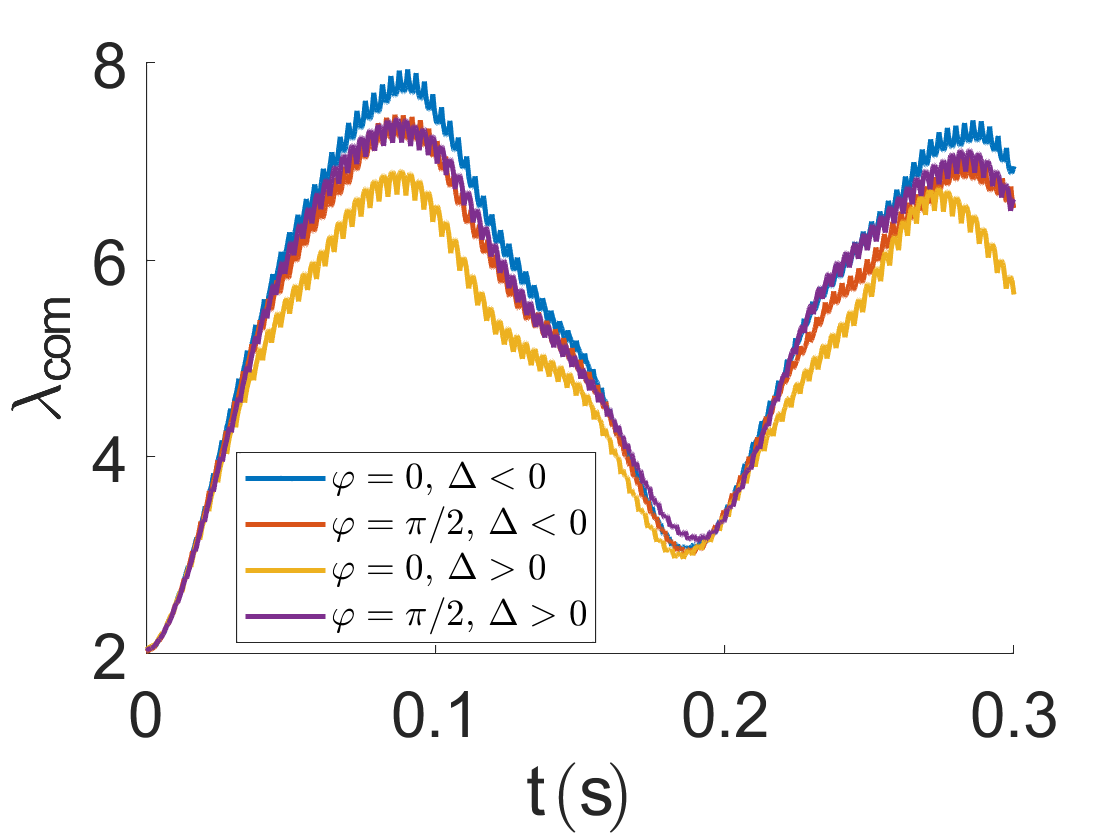}
	\caption{The effect on the synthetic space dynamics of different signs of the detuning for the two initial driving phases used for all the numerics. For $\varphi = 0$ (\textit{blue} and \textit{yellow}), we see a small difference in oscillation amplitude of around 1 synthetic lattice site for different signs of $\Delta$, whereas for $\varphi = \pi / 2$ (\textit{red} and \textit{purple}) the amplitudes are unchanged. We use the same parameters as Fig.~\ref{fig:da_unprocessed}, but with $|\Delta| = 5 \times 2\pi$ Hz.}
	\label{fig:det_sign}
\end{figure}
As an aside, we mention the effect of different signs of detuning (i.e. different signs of the effective force) on the dynamics in synthetic space. In Fig.~\ref{fig:det_sign}, we plot the synthetic space dynamics for both $\varphi = 0$ (\textit{blue} and \textit{yellow}) and $\pi / 2$ (\textit{red} and \textit{purple}) for opposite detuning signs. We see that, for $\varphi = 0$, there is a small amplitude difference of around 1 synthetic lattice site, whereas for $\varphi = \pi / 2$ the two amplitudes are the same. This means that we expect our oscillation amplitudes to be similar regardless of the sign of $\Delta$, as we find in our results in the main text. Note that the presence of a hard wall boundary at $\lambda = 0$ makes this different to the expected result for a wavepacket prepared in the synthetic dimension bulk. In that case, we expect the wavepacket to move in opposite directions for opposite signs of $\Delta$.

Secondly, we need to take into account the time-of-flight expansion (TOF) carried out in the experiment, as this is not included in the numerical method described in Appendix C. If this is not accounted for, then the experimental COM oscillations will be significantly larger than in the simulation. To correct for this in the numerics, we use the simulated momentum distribution to calculate the COM momentum in real space, $p_{\text{com}}$, and hence find the semiclassical cloud COM velocity $v_{\text{com}} = p_{\text{com}}/m$ at each time-step, such that we can correct the COM position from the numerics as:
\begin{equation}
    \tilde{x}_{\text{com}} = x_{\text{com}} + v_{\text{com}}t_{\text{tof}},
\end{equation}
where $t_{\text{tof}}$ is the TOF expansion time, corresponding to 5ms in this experiment. This method applied to our example oscillation (Fig.~\ref{fig:da_unprocessed}) is shown in Fig.~\ref{fig:analysis_steps}(a). We see the same qualitative form as Fig.~\ref{fig:da_unprocessed}(b), but with an amplitude around five times larger.
\begin{figure}[t!]
    \centering
    \includegraphics[width=0.45\textwidth]{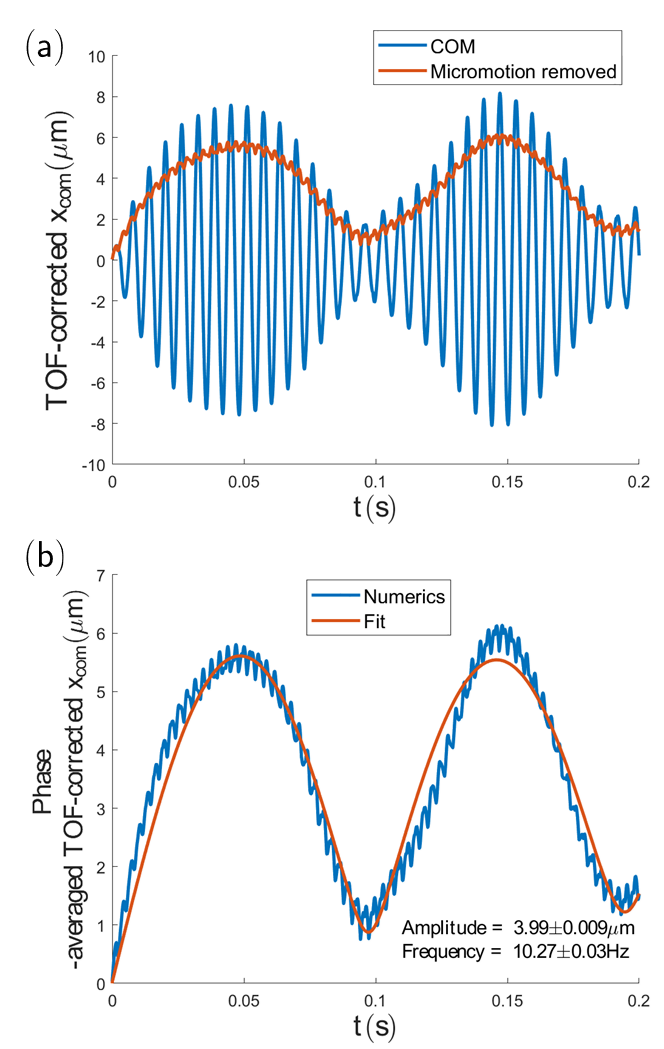}
    \caption{Steps in the analysis of simulated real space COM data. (a): 
    example real space COM with an applied TOF correction and $\varphi = 0$, showing a larger-magnitude oscillation (\textit{blue}), together with its RMS average over driving phases (\textit{red}), with $\varphi = 0, \pi/2$. The RMS-averaged curve shows a smaller amplitude than the full oscillation. (b): fit of Eq.~\ref{eqn:fit_app} (\textit{red}) to the TOF-corrected, RMS-averaged example oscillation (\textit{blue}), showing that our fitting function captures the dynamics well. The error bars on the reported fitted amplitude and frequency are those provided by the least-squares fit. These results use identical parameters to Fig.~\ref{fig:da_unprocessed}, and to Fig. 1(c) and 3 in the main text.}
    \label{fig:analysis_steps}
\end{figure}
Thirdly, as discussed above and in the main text, we need to remove the micromotion before we can extract the amplitude and frequency of the Bloch oscillation. Experimentally, this is done by repeating the experiment for multiple different starting phases $\varphi$ of the driving potential, as discussed in Appendix B. The micromotion is approximately removed by taking the root-mean-square (RMS) average over the $M$ chosen driving phases:
\begin{equation}
    \langle \tilde{x}_{\text{com}}\rangle = \sqrt{\frac{1}{M}\sum_{i = 1}^M (\tilde{x}_{\text{com}}^{(i)})^2}, \label{eqn:average}
\end{equation}
where $\tilde{x}_{\text{com}}^{(i)}$ is the COM for the $i$-th initial driving phase. Physically, the choice of $\varphi$ controls the phase of the micromotion oscillations. This approach is applied directly to the experimental data, as well as to the simulated data after the TOF correction. 

If the driving phases are chosen randomly or if the micromotion is very complicated, then we expect to take the large $M$ limit in Eq.~\ref{eqn:average} and average over many experimental or numerical runs. However, if the micromotion were described by a perfectly sinusoidal function, such as e.g. $ f(t) = \sin \omega_D t$, then we would in fact only need to average over any two values of $\varphi$ that are separated by $(2k + 1)\pi / 2$, with  $k = 0, 1, 2 ...$, using the property that $(f(t))^2 + (f(t+(2 k + 1 ) \pi/2))^2 = 1$. 

In practice, our numerical simulations (Fig.~\ref{fig:analysis_steps}(a) and~\ref{fig:da_unprocessed}) suggest that the micromotion is close to being sinusoidal and so we try averaging over only one pair of phases related by $(2k + 1)\pi / 2$. Indeed, Fig.~\ref{fig:analysis_steps}(a) shows an example of this RMS-averaged oscillation (\textit{red curve}) over two phases ($\varphi = 0, \pi/2$), together with the $\varphi = 0$ un-averaged data (\textit{blue curve}). As can be seen, averaging over only two runs is already sufficient to remove the majority of the micromotion, with a small residual that could be removed by using more pairs of phases. However, the amplitude of the averaged curve is smaller than the un-averaged result, and we will return to this point later when discussing corrections to the analytical results in Appendix G.
\begin{figure}
     \centering
     \includegraphics[width=0.45\textwidth]{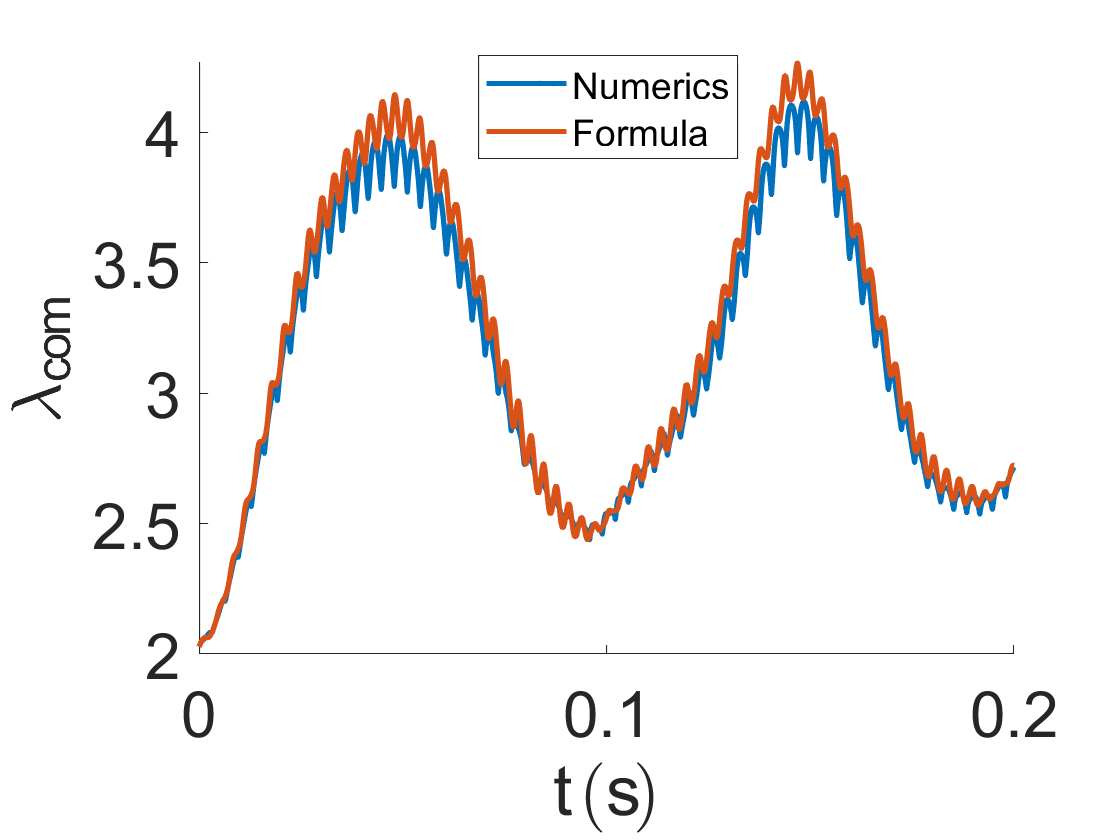}
     \caption{Example of the conversion formula (Eq.~\ref{eqn:conversion_formula_sm}, \textit{red}) applied to the same numerical simulation as Fig.~\ref{fig:da_unprocessed} (\textit{blue}), showing agreement up to a small offset. The formula is applied using the RMS time-averaged data in Fig.~\ref{fig:da_unprocessed}(b) and (c).}
     \label{fig:lambda_formula_example}
\end{figure}
Finally, to extract the oscillation amplitude and frequency, we fit the function:
\begin{equation}
    x(t) = A\sqrt{1 - e^{-gt}\cos(2\pi ft + \phi)},
    \label{eqn:fit_app}
\end{equation}
to the real space COM in both simulation and experiment, as introduced in the main text. This functional form is motivated by the analytical results; if we convert the expected oscillation in synthetic space (Eq. 4 in the main text) to real space (using Eq. 6 in the main text which is derived in Appendix E below), we obtain:
\begin{equation}
    x_{\text{com}} = \sqrt{\lambda_{\text{com}}^0 - \sigma_x^2 +\frac{1}{2} + \frac{2J}{\hbar\Delta}(1 - \cos(\Delta t))}.
\end{equation}
Since, in our numerics, $x_{\text{com}}(t = 0) = 0$, we have $\lambda_{\text{com}}^0 - \sigma_x^2 + 1/2 = 0$, where we assume that $\sigma_x$ is constant in time as justified previously. This then suggests the functional form of the above fitting function (Eq.~\ref{eqn:fit_app}). By hand, we then add in the exponential damping factor, to capture wavepacket splitting effects (see below) and other sources of damping in experiment (e.g. heating), as well as the phase $\phi$ to account for random variation in the position of the cloud after state preparation in the experiment. (Note that for  the numerics we have defined $x(t = 0) = 0$, and so we set $\phi = 0$ to reduce the number of fitting parameters.) Figure~\ref{fig:analysis_steps}(b) shows an example fit (\textit{red curve}) using Eq.~\ref{eqn:fit_app} applied to our numerical data (\textit{blue curve}) from Fig.~\ref{fig:analysis_steps}(a). The error bars on the fit parameters are from the least-squares fit. The fitted amplitude $A$ and frequency $f$ are plotted in the main text for varying detuning and shaking power.

\section*{Appendix E: Mapping from the synthetic dimension to real space}

Analytical results for our synthetic dimension scheme are expressed in terms of the synthetic dimension $\lambda$, but the experiment naturally probes real space $x$. We therefore now derive a formula linking these, which is given as Eq.~6 in the Main Text, and which is used to justify the fitting function as discussed above, and to process the analytical results in Appendix G. 
Our aim is to link the COM of a state $\ket{\psi}$ with respect to the synthetic dimension, $\lambda_{\text{com}}$, to the COM and width of the state in real space, $x_{\text{com}}$ and $\sigma_x$ respectively. To start, we will expand our state in the harmonic-trap basis as:
\begin{equation}
    \ket{\psi} = \sum_{\lambda}c_{\lambda}\ket{\lambda},
    \label{eqn:lambda_expansion}
\end{equation}
where $c_{\lambda}$ are complex coefficients with $\sum_{\lambda}|c_{\lambda}|^2 = 1$. In terms of the COM variables, it is straightforward to show that
\begin{equation}
    \lambda_{\text{com}} = |\alpha|^2, \text{ } \alpha \equiv \frac{x_{\text{com}} + ip_{\text{com}}}{\sqrt{2}},
\end{equation}
where: 
\begin{align}
    x_{\text{com}} = \bra{\psi}\hat{x}\ket{\psi}, \\
    p_{\text{com}} = \bra{\psi}\hat{p}\ket{\psi},
\end{align}
are the COM in real and momentum space respectively. Here, $x_{\text{com}}$ is measured in units of $\sqrt{\hbar / m\omega_x}$, and $p_{\text{com}}$ in units of $\sqrt{\hbar m\omega_x}$. We can also write
\begin{equation}
    \sigma_x^2 = \bra{\psi}\hat{x}^2\ket{\psi} - x_{\text{com}}^2,
    \label{eqn:width}
\end{equation}
which follows from the usual expression for the variance of a random variable. Now writing $\hat{x} = \frac{1}{\sqrt{2}}(\hat{a} + \hat{a}^{\dagger})$, where $\hat{a}^{\dagger}$ and $\hat{a}$ are the usual harmonic oscillator raising and lowering operators, and substituting into Eq.~\ref{eqn:width}, yields:
\begin{equation}
    \sigma_x^2 = \bra{\psi}\left(\hat{a}^{\dagger}\hat{a} + \frac{1}{2}\right)\ket{\psi} + \frac{1}{2}(\bra{\psi}\hat{a}^2\ket{\psi} + \bra{\psi}\hat{a}^{\dagger 2}\ket{\psi}) - x_{\text{com}}^2,
\end{equation}
where we have used the commutator $[\hat{a}, \hat{a}^{\dagger}] = 1$. Now inserting our expansion of $\ket{\psi}$ (Eq.~\ref{eqn:lambda_expansion}) produces:
\begin{equation}
  \sigma_x^2 = \lambda_{\text{com}} + \frac{1}{2} +\frac{1}{2}(S + S^{*}) - x_{\text{com}}^2, 
\end{equation}
where we identified $\lambda_{\text{com}} = \sum_{\lambda}|c_{\lambda}|^2\lambda$ and where:
\begin{equation}
    S = \sum_{\lambda = 0}^{\infty}c_{\lambda}^*c_{\lambda + 2}\sqrt{(\lambda + 1)(\lambda + 2)}.
\end{equation}
Finally, re-arranging for $\lambda_{\text{com}}$ gives:
\begin{equation}
    \lambda_{\text{com}} = x_{\text{com}}^2 + \sigma_x^2 - \frac{1}{2} - \frac{1}{2}(S + S^*).
\end{equation}
To gain some intuition for the $S$ terms, consider the limit of preparing the system in a particular eigenstate of the harmonic trap $\ket{\lambda_0}$, so $c_{\lambda} = \delta_{\lambda, \lambda_0}$. This makes $S = 0$, and we have:
\begin{equation}
    \lambda_{\text{com}} = x_{\text{com}}^2 + \sigma_x^2 - \frac{1}{2}.
    \label{eqn:conversion_formula_sm}
\end{equation}
If, instead, the state is a semiclassical Gaussian wavepacket with $c_{\lambda} \sim \exp({-(\lambda - \lambda_0)^2/2\sigma_{\lambda}^2})$, we expect Eq.~\ref{eqn:conversion_formula_sm} to hold approximately when $\sigma_{\lambda}$ is sufficiently small. Figure~\ref{fig:lambda_formula_example} shows an example of this result (\textit{red curve}) compared against the COM calculated directly from a numerical simulation (\textit{blue curve}), with agreement up to a small offset. The numerical curve displays small micromotion oscillations as expected, as does the curve calculated from our formula. The formula shows these oscillations because using only a single pair of phases during micromotion averaging does not perfectly remove the micromotion, as discussed in Appendix D. Note also that the example data and parameters used in this section are the same as the previous one, although the level of agreement is similar in all cases studied. This result demonstrates that the derived conversion formula still holds in the case of the thermal cloud. 
\begin{figure}[t!]
	\includegraphics[width=0.45\textwidth]{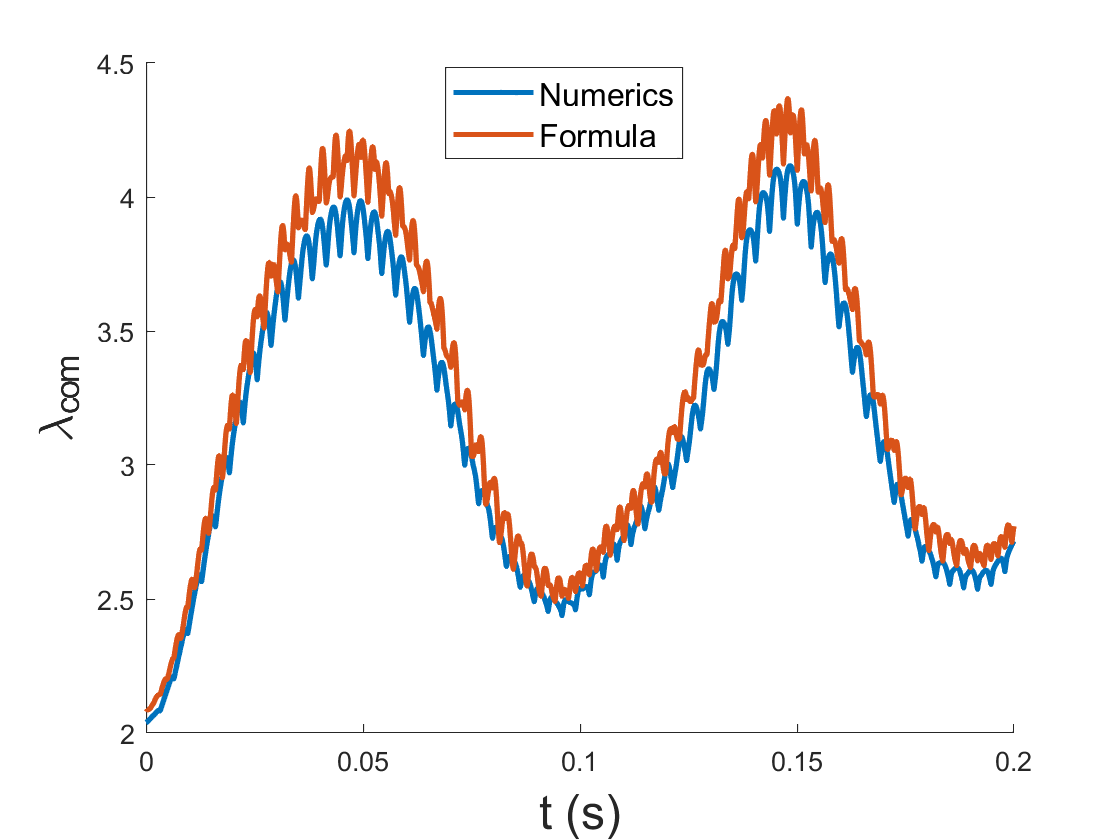}
	\caption{The momentum space conversion formula (Eq.~\ref{eqn:conversion_formula_momentum}) applied to the same oscillation as Fig.~\ref{fig:lambda_formula_example}. The result of applying Eq.~\ref{eqn:conversion_formula_momentum} to the RMS phase-averaged momentum COM and width (\textit{blue}) is compared to the synthetic space COM calculated numerically (\textit{red}) and we see agreement up to a small offset.}
	\label{fig:lambda_formula_example_p}
\end{figure}
\begin{figure*}
	\includegraphics[width=\textwidth]{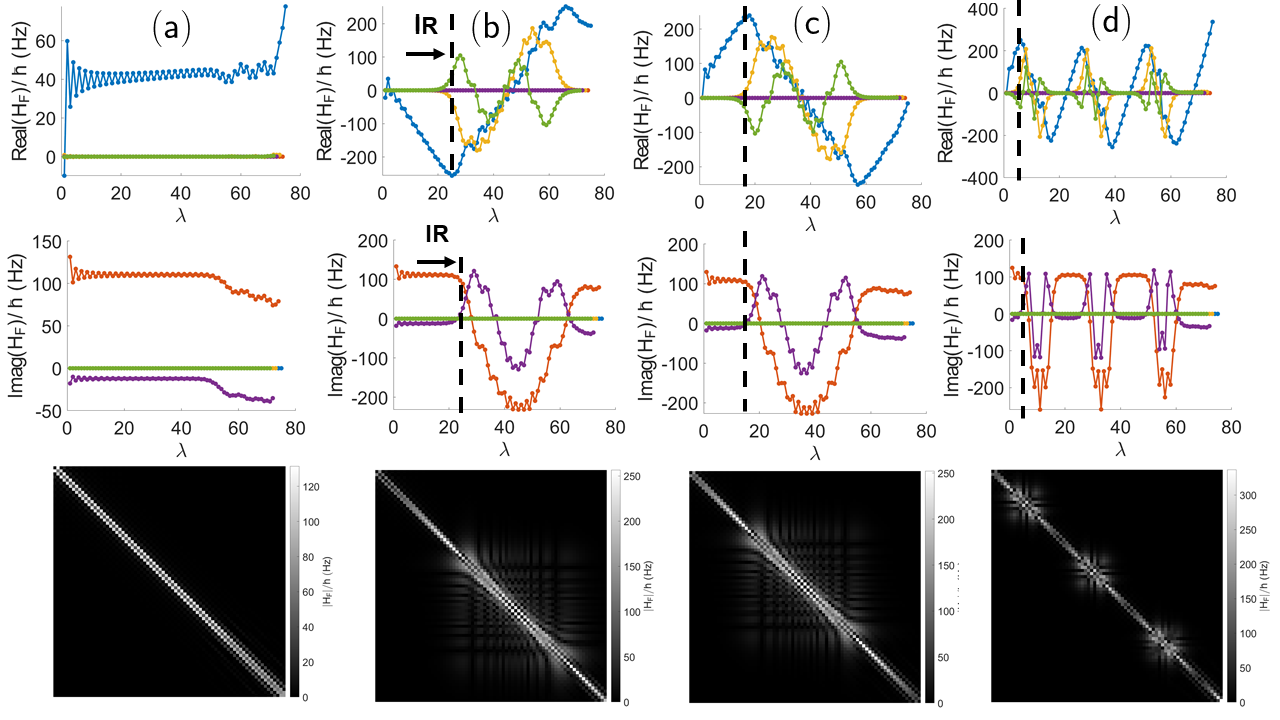}
	\caption{Numerically-calculated matrix elements of the Floquet Hamiltonian (Eq.~\ref{eqn:HF}) for $\omega_x = 2\pi \times 166.5 \text{ Hz}$, $V_0 = 4.16 \text{ nK}$ and $\varphi = 0$, for $\Delta = 2\pi \times 0, -2, 2 \text{ and } 7 \text{ Hz}$ in columns (a), (b), (c) and (d) respectively. These parameters correspond to the low-power data shown in Fig. 1, 2, 3 and 4 in the main text. The top row shows the real part of the first five diagonals, the middle row shows the imaginary part, and the final row shows a heat map of $|\hat{H}_F|$. The on-site terms are in blue, the NN hoppings in red and longer-range hoppings in other colours. We see that we can approximate $\hat{H}_F$ by a nearest-neighbour tight-binding model where $\Delta$ plays the role of a force along the synthetic dimension. The onset of instability regions (IR), marked by the black dotted lines, show long-range hoppings for some $\lambda$ and are a numerical artifact and are not physical.}
	\label{fig:me_low_power}
\end{figure*}
We can repeat the above calculation but working in terms of momentum rather than position to find:
\begin{equation}
    \lambda_{\text{com}} = p_{\text{com}}^2 + \sigma_p^2 - \frac{1}{2},
    \label{eqn:conversion_formula_momentum}
\end{equation}
where $\sigma_p$ is the width of the state in momentum space, and the momenta are measured in units of $\sqrt{\hbar m\omega_x}$. An example of this formula applied to the oscillation in Fig.~\ref{fig:lambda_formula_example} is shown in Fig.~\ref{fig:lambda_formula_example_p}, showing similar agreement to the real space result. 

\section*{Appendix F: Effective Time-Independent Hamiltonian Description}
\begin{figure}
	\includegraphics[width=0.45\textwidth]{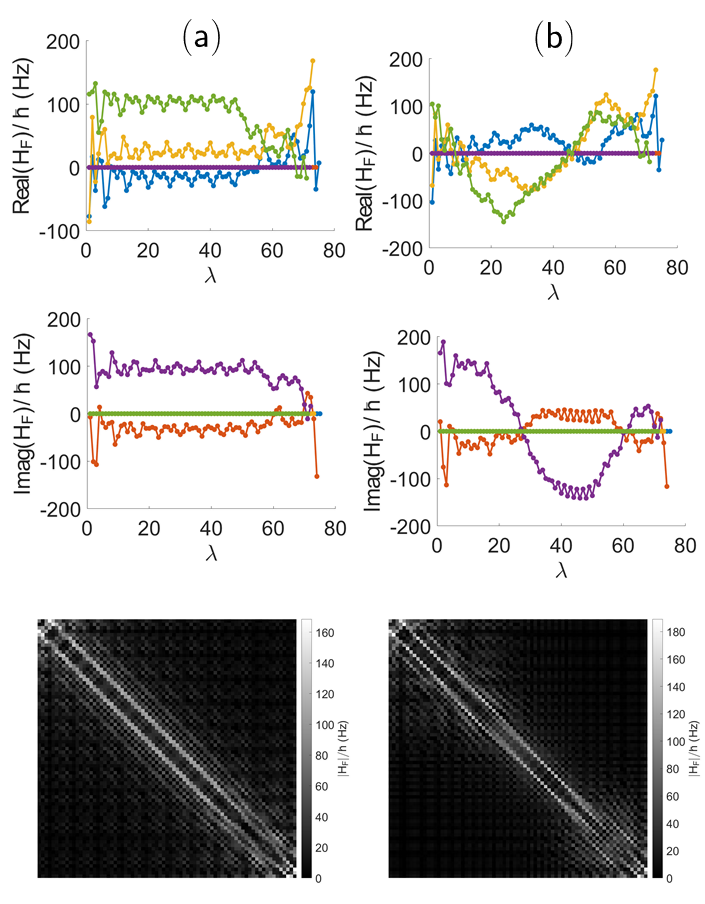}
	\caption{Floquet Hamiltonian matrix elements for a higher shaking power than Fig.~\ref{fig:me_low_power}. We have $\omega_x = 2\pi \times 142.1 \text{ Hz}$, $V_0 = 11.96 \text{ nK}$ and $\varphi = 0$, and $\Delta = 2\pi \times 0$ and $-2 \text{ Hz}$ in columns (a) and (b) respectively. These parameters correspond to the high-power data shown in Fig. 3(a) in the main text. Unlike the low-power case, we find significant long-range hoppings for all $\lambda$, and that a detuning does not simply induce a slope on the on-site terms. This is not a numerical artifact, but a result of the failure of the rotating-wave approximation, and precludes building a simple effective Hamiltonian. We use the same layout and colour schemes as Fig.~\ref{fig:me_low_power}.}
	\label{fig:me_high_power}
\end{figure}
In this section, we show how we arrive at the effective time-independent Hamiltonian shown in the main text:
\begin{equation}
    \hat{\mathcal{H}} = \hbar\Delta \sum_{\lambda} \lambda \ket{\lambda}\bra{\lambda}  + J \sum_{\lambda} \left[ ie^{ i \varphi} \ket{\lambda + 1}\bra{\lambda} + \text{h.c} \right] \label{eqn:H_app}
\end{equation} 
starting from the full time-dependent Hamiltonian (Eq. 2 in the main text). Note that this effective Hamiltonian is different to that of~\cite{Price2017} in the sense that we can assume that our nearest-neighbour hoppings are constant in $\lambda$, whereas those of~\cite{Price2017} scale as $\sqrt{\lambda}$; this is due to the choice of driving potential. 
As the Hamiltonian is periodic in time, we can use Floquet theory to define the Floquet Hamiltonian, $\hat{H}_F$, according to:
\begin{equation}
    \hat{H}_F = \frac{i\hbar}{T_D} \log( \hat{U} (T_D; 0))
\label{eqn:HF}
\end{equation}
where $\hat{U} (T_D; 0) $ is the time-evolution operator over a full period of the driving, $T_D = 2 \pi / \omega_D$, where $\omega_D$ is the driving frequency~\cite{Goldman2014}. This Hamiltonian can be calculated numerically by splitting the time-evolution operator (Eq.~\ref{eqn:HF}) into sufficiently many small timesteps $dt$. For simplicity, we have neglected terms in the 
Hamiltonian that depend on the $y$ - direction to work with a 1D Hamiltonian in the $\lambda$ - basis.

We first investigate the matrix elements of Eq.~\ref{eqn:HF} for the low-$V_0$ case ($V_0 = 4.16 \text{ nK}$), as considered in Fig. 1, 2, 3 and 4 in the main text. The Floquet Hamiltonian matrix elements are plotted in Fig.~\ref{fig:me_low_power} for several detunings and for parameters listed in the caption. More precisely, we plot the real (top row) and imaginary (middle row) parts of the first five diagonals of $\hat{H}_F$, as well as $|\hat{H}_F|$ as a heat map to show the longer-range structure of the Hamiltonian (bottom row). For all figures in this section, the on-site terms are in blue, the nearest-neighbour (NN) hoppings are in red, and longer-range hoppings in other colours. For this low-power figure, we have applied a constant offset to the time-dependent Hamiltonian to ensure that the on-site matrix elements are zero for $\lambda = 0$. This shifts the location of the instability regions (see below) and allows the behaviour for $\Delta < 0$ to be seen more clearly.

As can be seen, for many values of $\lambda$, the Hamiltonian can be approximated by the form used in the main text. Firstly, the detuning induces a tilt on the on-site matrix elements (\textit{blue curve}) $\sim \Delta\lambda$, which allows us to interpret this detuning as a constant force along the synthetic dimension (see e.g. Fig.~\ref{fig:me_low_power} columns (b), (c) between $\lambda \approx 0$ and $\lambda \approx 20$). Note that we verified that the slope of the on-site terms is equal to the detuning by fitting a straight line to these plots. Secondly, across the same regions, we also find nearly-flat NN hoppings (\textit{red curve}), such that the NN hopping energy $J$ can be calculated by taking the average of the NN matrix elements up to the onset of the instability region (see below). This hopping energy does not vary significantly as a function of detuning, and we find the value $J/\hbar = 106 \pm 8 \text{Hz}$ for $V_0 = 4.16\text{nK}$. To calculate this error bar on $J$, we add the standard deviations of the NN hoppings for each of the $N_{\Delta}$ detunings in quadrature to obtain $\sigma$, and then use the error bar $\sigma / \sqrt{N_{\Delta}}$. Thirdly, we also see that we have a small long-range hopping (\textit{purple curve}) which we neglect. Note that all the matrix elements show small oscillations with respect to $\lambda$, which may act as a potential minimum around $\lambda = 0$ and be the cause of the cloud splitting effects discussed in the main text and below. This is discussed further in Appendix I. Finally, we see the ``odd'' diagonals of the Hamiltonian (nearest-neighbour, NNNN, etc.) are purely imaginary and the others are purely real, which is caused by the initial driving phase $\varphi$ playing the role of a Peierls phase in the effective model.

For non-zero detuning in Fig.~\ref{fig:me_low_power}, there also appear to be regions of $\lambda$ where we no longer have linear on-site terms and flat NN hoppings, but instead have significant long-range hoppings (e.g. between $\lambda \approx 20 $ and $\lambda \approx 60$ in column (b)). However, these are an artifact of our numerics and should not be physical. These regions arise because the Floquet Hamiltonian is not unique, but depends on the branch of the matrix logarithm (Eq.~\ref{eqn:HF}). In the numerics, the principal branch is always taken, meaning that the Floquet Hamiltonian is constructed to have eigenvalues that only lie between $-\pi / T_D$ and $\pi / T_D$. This leads to the apparent``wrapping around'' of onsite terms and an associated variation in off-diagonal terms when the on-site shift due to the detuning becomes large. This can be seen by noting that for larger detuning (column (d)), the apparent breakdown happens earlier and more frequently, and between these regions, the matrix elements look regular and well-behaved. We have also checked our interpretation numerically by adding a constant offset to the time-dependent Hamiltonian $\hat{H}(t)$ that changes the size and location of the apparent breakdown regions while leaving the on-site slope and NN hoppings otherwise unchanged. Finally, we also do not observe any qualitative change in behaviour of our numerical simulations (Appendix C) when the cloud moves in the instability regions, further confirming that these are not a physical effect.

It is important to distinguish between the above numerical artifact and a genuine breakdown of the effective Hamiltonian employed in the low-$V_0$ case, caused by the failure of the rotating-wave approximation. This is apparent e.g. in Fig.~\ref{fig:me_low_power} (a) above $\lambda \approx 55$, where we observe that the matrix elements begin to deviate from their previous values.

For larger values of $V_0$ (such as considered in Fig. 2 in the main text), we also observe numerically that the matrix elements become less uniform and more long-ranged. This is shown in Figure~\ref{fig:me_high_power}, where we plot the Floquet Hamiltonian matrix elements for $V_0 = 11.96 \text{ nK}$ for two detunings, with parameters as listed in the figure caption. As can be seen, in this case, we find significant long-range hoppings for  all $\lambda$, and a non-zero detuning does not simply add a slope to the on-site terms.  This therefore precludes building a simple analytical model for this behaviour, although we still find that our numerical simulations agree well with experimental results, as in Fig. 3 in the main text.

\section*{Appendix G: Details of Analytical Results}

Here we describe in detail how the analytical results for Bloch oscillations in both real and synthetic space are obtained, including corrections to make them comparable to numerical and experimental data, as plotted and discussed in the main text.

As stated in the main text, we expect the cloud COM to oscillate with respect to the synthetic dimension as:
\begin{equation}
    \lambda_{\text{com}}(t) = \lambda_{\text{com}}^0 + \frac{2J}{\hbar\Delta}(1 - \cos(\Delta t)).
    \label{eqn:lambda_sm}
\end{equation}
We therefore expect an oscillation frequency of $\Delta / 2\pi$ and a maximum $\lambda$ of $\lambda_{\text{max}} = \lambda_{\text{com}}^0 + 4J / \hbar \Delta$. We can then use our result connecting the real and synthetic dimension, Eq.~\ref{eqn:conversion_formula_sm}, to calculate the maximum displacement of the cloud from $x = 0$. This gives:
\begin{equation}
    x_{\text{max}} = \sqrt{\lambda_{\text{max}} - \sigma_x^2 + \frac{1}{2}},
    \label{eqn:real_space_basic}
\end{equation}
as discussed in the main text, and where we measure $x_{\text{max}}$ and $\sigma_x$ in units of $\sqrt{\hbar/m\omega_x}$. We can analytically calculate the width for a thermal cloud $\sigma_x = \sqrt{k_BT / m\omega_x^2}$~\cite{Pethwick2001}, and approximate the cloud width as constant in time, as justified in Appendix D. 

We now take into account cloud splitting, as described in the main text. Note that we discuss methods to reduce this effect in Appendix I. The COM with respect to the synthetic dimension is calculated as:
\begin{equation}
    \lambda_{\text{com}}(t) = \sum_{\lambda}\lambda\rho(\lambda, t),
\end{equation}
where $\rho(\lambda, t)$ is the synthetic space density, calculated numerically as in Section I, where we have integrated out the $y$-dependence. The presence of a split wavepacket component skews this average downward, and the analytical result (Eq.~\ref{eqn:lambda_sm}) therefore overestimates numerical and experimental amplitudes.

To correct for this, we define a cutoff in $\lambda$, $\lambda_c$, that cleanly separates the two wavepacket components near the peak of the oscillation. In particular, we choose $\lambda_c = 2J/ \hbar \Delta$, because this is about half of the maximum $\lambda$ coordinate at the oscillation peak (Eq.~\ref{eqn:lambda_sm}). Near the oscillation peak, we can then write: 
\begin{align}
    \lambda_{\text{com}} = (1 - r)\lambda_{\text{com}}^{<} + r\lambda_{\text{com}}^>, \\
    r = \sum_{\lambda > \lambda_c}\rho(\lambda),
\end{align}
where $\lambda_{\text{com}}^{<}$ ($\lambda_{\text{com}}^{>}$) is the centre of mass of the lower (upper) wavepacket respectively, and $r$ is the amount of wavepacket above the cutoff. We can calculate $r$ numerically and find that it is approximately constant with respect to the detuning, with an average value of $r = 0.52 \pm 0.03$, where the error bar is the standard deviation of the $r$ values over the detunings. This value is for the low-power data, with $V_0 = 4.16 \text{ nK}, \omega_x = 166.5 \times 2\pi \text{ Hz}, \omega_y = 10 \times 2\pi \text{ Hz}$ and $T = 20 \text{ nK}$. To make our analytical result for $\lambda_{\text{max}}$ comparable to the numerics and experiment, we hence correct it as:
\begin{equation}
    \lambda_{\text{max}} = \lambda_{\text{com}}^0 + r\frac{4J}{\hbar \Delta}.
\end{equation}
\begin{figure}
	\includegraphics[width=0.45\textwidth]{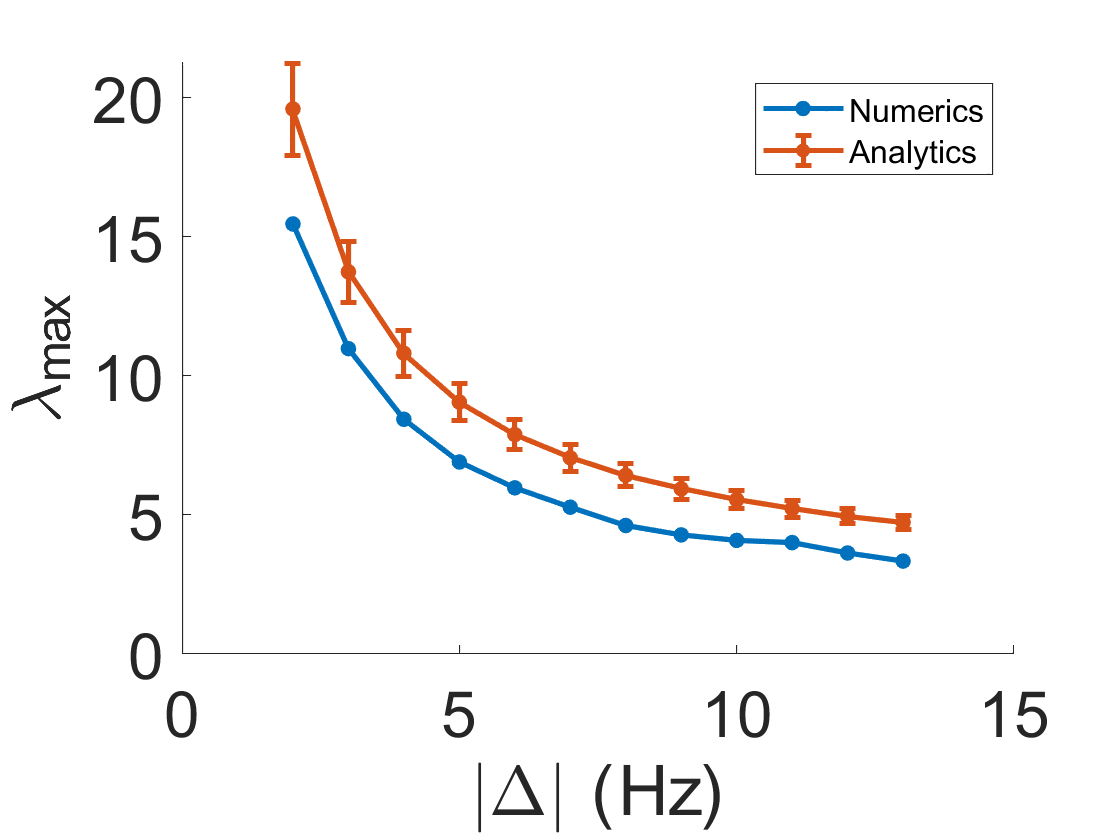}
	\caption{Cloud-splitting corrected analytical synthetic space amplitude (\textit{red}) compared to the maximum $\lambda_{\text{com}}$ obtained numerically (\textit{blue}). We see good agreement, up to a constant offset of around two synthetic lattice sites. Error bars on analytical results are calculated by propagating errors on numerical parameters. Lines are a guide to the eye. Here, we use parameters for the low-power data, as in, for example, Fig.~\ref{fig:analysis_steps}.}
	\label{fig:lambda_amplitude}
\end{figure}
This result is plotted in Fig.~\ref{fig:lambda_amplitude} (\textit{red}) and compared against the maximum $\lambda_{\text{com}}$ in the numerics (\textit{blue}). We see good agreement, up to a roughly constant offset of around two synthetic lattice sites. We can also use this corrected expression for $\lambda_{\text{max}}$ in our real space result (Eq.~\ref{eqn:real_space_basic}) to get: 
\begin{equation}
    x_{\text{max}} = \sqrt{\lambda_{\text{com}}^0 + r\frac{4J}{\hbar \Delta} - \sigma_x^2 + \frac{1}{2}}.
\end{equation}
We now need to apply three further corrections to this real space result in order to compare to the experimental and numerical results including TOF. 

Firstly, we need to estimate the effect of TOF expansion. To do this, we numerically calculate the ratio of the maximum of the real space oscillation including TOF (Appendix D) to the maximum of the oscillation without. We find that this is approximately independent of detuning, and has a value of $\alpha_{\text{tof}} = 5.25 \pm 0.08$, where the error bar is the standard deviation of the $\alpha_{\text{tof}}$ values over the detunings. As for $r$ above, this is calculated for the low-power data. We then correct our real space result as:
\begin{equation}
    x_{\text{max}} \to \alpha_{\text{tof}}x_{\text{max}}.
\end{equation}
\begin{figure}
	\includegraphics[width=0.45\textwidth]{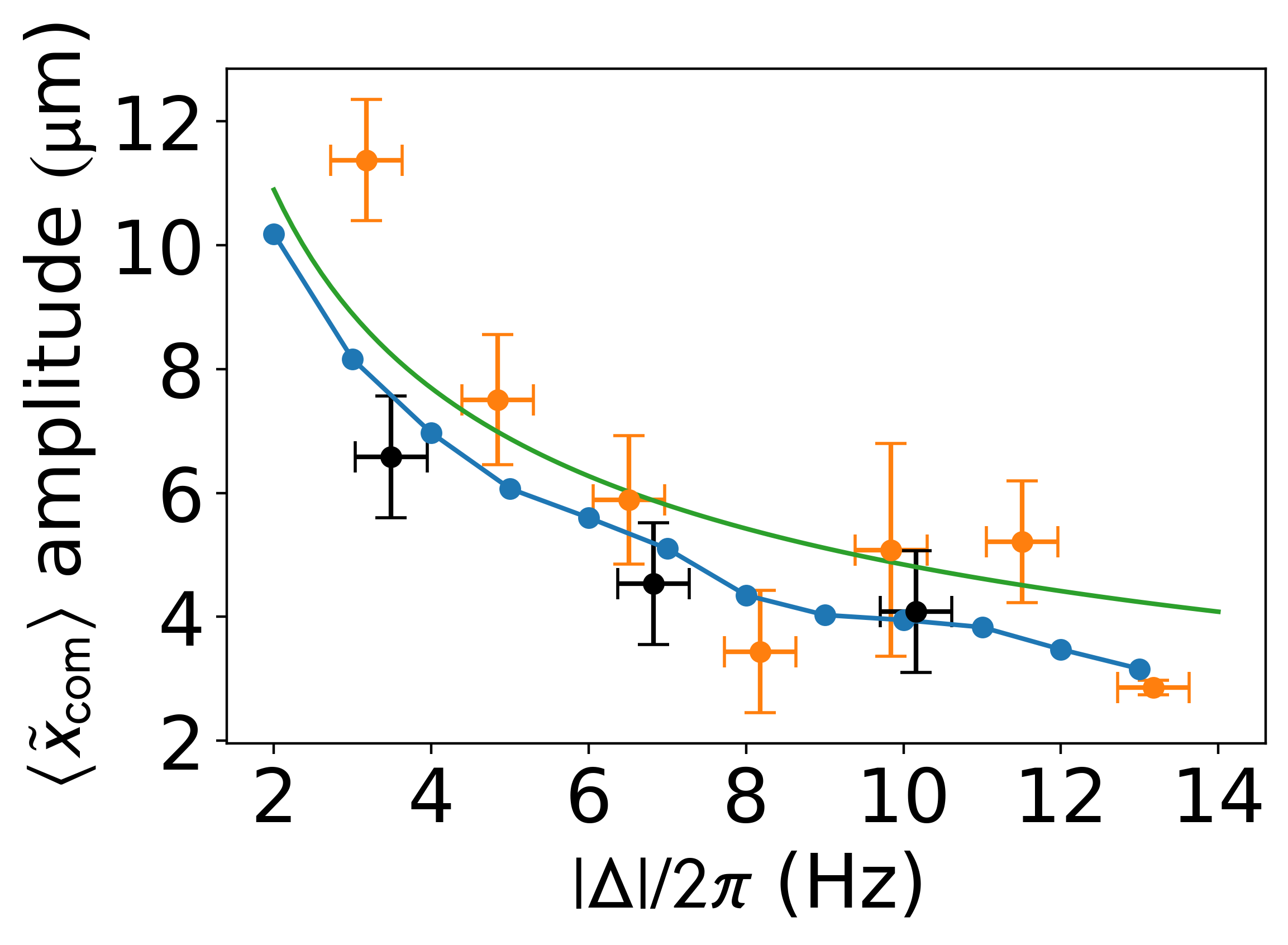}
	\caption{Bloch oscillation amplitude for variable detuning as in Fig. 2(b) in the main text. The experimental amplitude fit parameter (\textit{orange}) is compared to the fit parameter for the numerics (\textit{blue}) and the analytical result (\textit{green}), where the TOF correction in the analytics is analytically estimated (Eq.~\ref{eqn:vcom}). We see that this TOF correction also produces good agreement between the analytics, experiment and numerics, although it overestimates the size of TOF effects. Three experimental datapoints (\textit{black}) are have their driving frequency above the trap frequency, whereas all others are below. Here we have: $\omega_x = 2\pi \times 166.5 \text{ Hz}$, $V_0 = 4.16 \text{ nK}$, $T = 20 \text{ nK}$, $\omega_y = 2\pi \times 10 \text{ Hz}$ and $\varphi = 0, \pi/2$.}
	\label{fig:alt_tof}
\end{figure}
Alternatively, we can consider correcting our result analytically by calculating the COM velocity $v_{\text{com}}$ at the oscillation peak. To do this, we can use the momentum space version of the $\lambda$ formula (Eq.~\ref{eqn:conversion_formula_momentum}) to calculate $p_{\text{com}}$ at the oscillation peak, including the wavepacket splitting correction, and then calculating $v_{\text{com}} = p_{\text{com}}/m$, to yield:
\begin{equation}
    v_{\text{com}} = \sqrt{\frac{\hbar}{m\omega_x}}\frac{1}{m}\sqrt{\lambda_{\text{com}}^0 + r\frac{4J}{\hbar \Delta} + \frac{1}{2} - \sigma_p^2},
    \label{eqn:vcom}
\end{equation}
where $\sigma_p$ is calculated from the numerics and assumed constant in time, and we measure $\sigma_p$ in units of $\sqrt{\hbar m\omega_x}$. We can then apply this time-of-flight correction to $x_{\text{max}}$:
\begin{equation}
    x_{\text{max}} \to x_{\text{max}} + v_{\text{com}}t_{\text{tof}}.
\end{equation}
The effect of doing this, together with the two corrections described below, is shown in Fig.~\ref{fig:alt_tof} (\textit{green curve}), where the blue curve is the amplitude fit parameter for the numerical simulations, and the orange points are the fit parameter for the experimental data, as discussed in Appendix D and the main text, and where we use the same low-power parameters as Fig. 1, 2, 3 and 4 in the main text. We see that the level of agreement between these analytics and the numerics and experiment is comparable to the method using a numerical TOF correction, although this method clearly overestimates the effect of TOF. 

Next, we correct for the effect of RMS averaging. As described in Appendix D, the RMS average over initial driving phases to remove micromotion slightly underestimates the true size of the oscillation (Fig.~\ref{fig:analysis_steps}(a)). If we model the dynamics of the COM as a beat:
\begin{equation}
    x_{\text{com}}(t) = A\sin(\omega_Dt)\sin\left(\frac{\omega_{B}t}{2}\right),
\end{equation}
where $\omega_{B}$ is the slow Bloch oscillation frequency, we can then RMS-average out the micromotion using two initial driving phases of 0 and $\pi / 2$:
\begin{multline}
    \langle x_{\text{com}}\rangle^2 = \frac{1}{2}((A\sin(\omega_Dt)\sin\left(\omega_{B}t / 2\right))^2 + \\ (A\sin(\omega_Dt + \pi / 2)\sin\left(\omega_{B}t / 2\right))^2).
\end{multline}
\begin{figure}[!t]
	\includegraphics[width=0.48\textwidth]{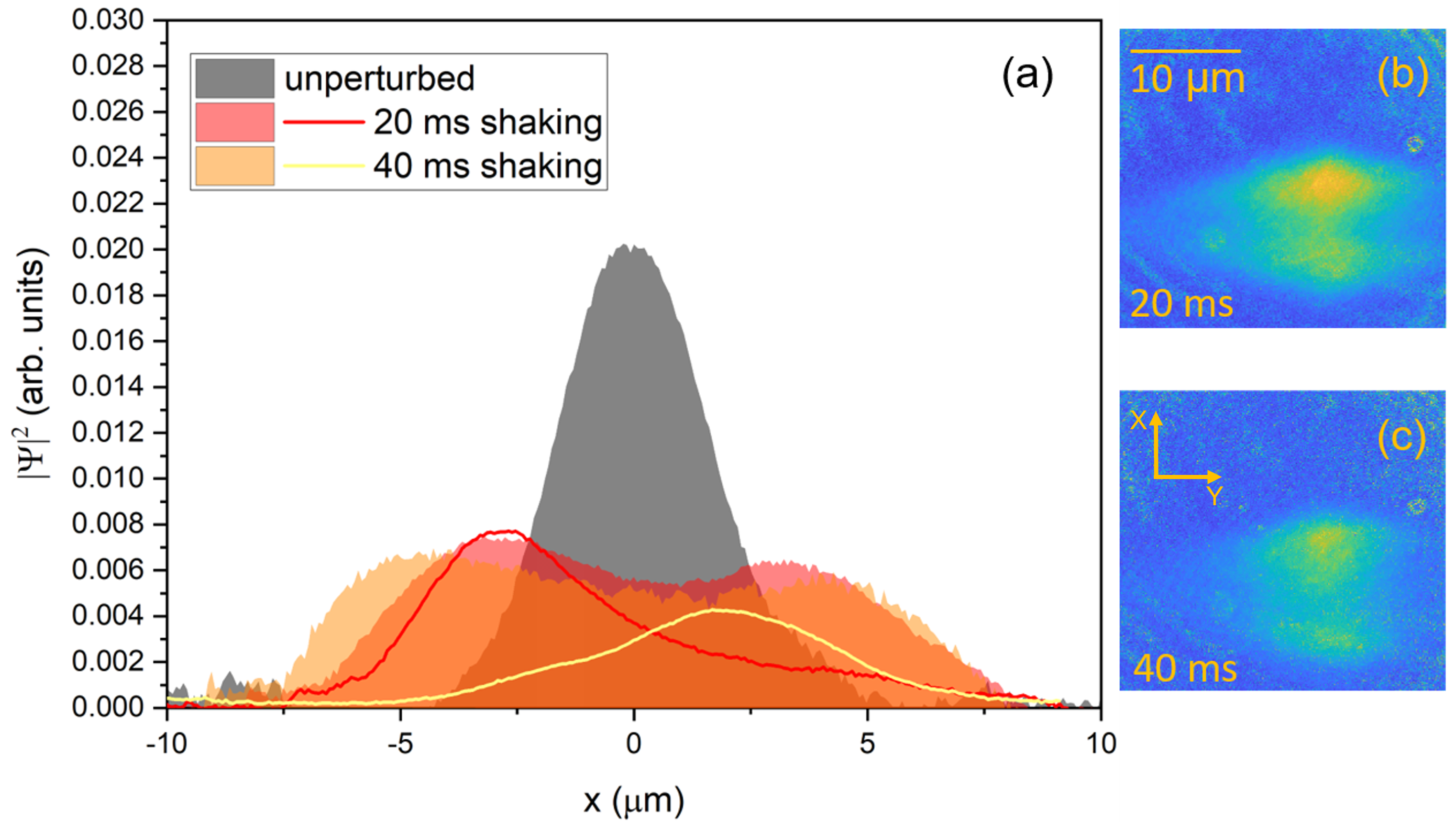}
	\caption{(a) Column density distribution integrated along the $y$ direction of our atomic samples after 5 ms of time of flight. The shaded areas correspond to the average of 30 experimental runs where we vary the shaking phase. The solid lines are examples of density distributions that are obtained with fixed phase. The number of atoms is $\simeq15\times10^4$. (b) and (c) are the column densities averaged over 30 runs with different phases in the shaking potential that we observe after 5 ms of time of flight and with 20 and 40 ms of shaking time respectively.}
	\label{fig:density}
\end{figure}

Here, we have shifted the micromotion part of the beat by $\pi / 2$. Simplifying gives $\langle x_{\text{com}} \rangle^2 = A^2\sin(\omega_Bt/2)^2/2$. We can then calculate a scale factor between the averaged and un-averaged results, $\alpha_{\text{rms}} \equiv \langle x_{\text{com}}\rangle_{\text{max}}/A = 1/\sqrt{2}$. We then apply the correction:
\begin{equation}
    x_{\text{com}} \to \alpha_{\text{rms}}x_{\text{com}}.
\end{equation}
Note that we can calculate a scale factor based on an infinite number of driving phases by doing the RMS time-average, and we find the same result. As discussed in Appendix D, our numerical results can be well-described by such a sinusoidal beat, so this approach works well for our data. 

Our final correction is due to the fitting function used (Eq.~\ref{eqn:fit_app}). Neglecting damping, our fitting function will have a maximum value of $x_{\text{max}} = \sqrt{2}A$, and we analytically calculate $x_{\text{max}}$. We therefore correct our analytical result so far as:
\begin{equation}
    x_{\text{max}} \to \frac{1}{\sqrt{2}}x_{\text{max}},
\end{equation}
so that we can compare directly to the fit parameters obtained from the numerics and experiment. Our final analytical result for $x_{\text{max}}$, as shown in the main text, is then:
\begin{equation}
    x_{\text{max}} = \frac{1}{\sqrt{2}}\alpha_{\text{rms}}\alpha_{\text{tof}}\left(\sqrt{\frac{\hbar}{m\omega_x}\left(\lambda_{\text{com}}^0 + r\frac{4J}{\hbar \Delta} + \frac{1}{2}\right) - \sigma_x^2}\right),
\end{equation}
where we have restored SI units. Finally, note that these same analytical results are used for the variable-power data in Fig. 3(b) in the main text, where the numerical parameters $\alpha_{\text{tof}}$, $J$ and $r$ were re-calculated for this dataset.

\begin{figure*}[!t]
	\includegraphics[width=\textwidth]{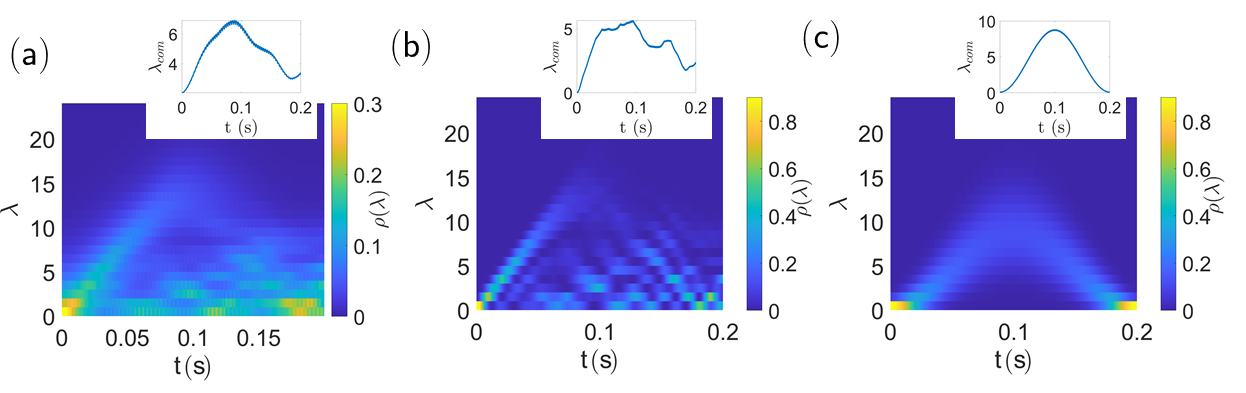}
	\caption{(a): Time evolution of the integrated synthetic space density for a thermal cloud under our usual driving potential using typical parameters, for a detuning of $\Delta = 5 \times 2\pi$ Hz. We see significant cloud splitting and a relatively wide oscillating component. We use identical parameters to our low-power data elsewhere in this work, with $\varphi = 0$. (b): The result in (a) but with increased trapping frequency and decreased temperature, showing that the  oscillating part of the cloud becomes much narrower in synthetic space. Note the different colour scale to (a). We use the same parameters as (a) but with $T = 5$ nK and $\omega_x = 2\pi \times 500$ Hz. (c): Time evolution of the integrated synthetic space density of a thermal cloud for a modified driving potential, showing that the entire cloud moves, with no split component. Note that the linear driving potential together with parameters as in (a) produces a wide oscillating cloud, so we also increase the trapping frequency and decrease the temperature here to compensate for this. We use: $\kappa x_{\text{qho}} = 1$ nK, with $x_{\text{qho}} = \sqrt{\hbar / m\omega_x}$, $\omega_x = 500 \times 2\pi$ Hz, $\Delta = 5 \times 2\pi$ Hz, $T = 10$ nK and $\omega_y = 10 \times 2\pi$ Hz. The insets of each panel shows the corresponding $\lambda$ COM as a function of time. Note that the results for (a) and (b) are not rescaled to take into account cloud splitting.}
	\label{fig:state_fidelity}
\end{figure*}

\section*{Appendix H: Synthetic dimension experimental data analysis and theoretical results}

To convert our real space experimental oscillation into synthetic space (Fig.~\ref{fig:F3}(a)), we use our conversion formula (Eq.~\ref{eqn:conversion_formula}), assuming that the real space width $\sigma_x$ (measured in harmonic oscillator lengths) remains constant in time and set by the trap frequency and initial temperature. We then obtain:
\begin{equation}
    \lambda_{\text{com}}(t) = \left(\frac{x_{\text{com}}(t)}{\alpha_{\text{rms}}\alpha_{tof}}\right)^2 + \sigma_x^2 - \frac{1}{2},
\end{equation}
where $x_{\text{com}}(t)$ is the experimental real space COM measured in harmonic oscillator lengths. The conversion for the data in Fig.~\ref{fig:F4} is similar, but here we convert the experimental amplitude fitting parameters, so we use the conversion: 
\begin{equation}
    \lambda_{\text{com}} = \left(\left(\frac{x_{\text{com}}}{\alpha_{\text{rms}}\alpha_{tof}/\sqrt{2}}\right)^2 + \sigma_x^2 - \frac{1}{2}\right)/r,
\end{equation}
This is then compared against the corresponding numerics, where we extract the maximum value of $\lambda_{\text{com}}$ and rescale by $1 / r$ to account for the cloud splitting. We also compare against an analytical expression, $\lambda_{\text{com}} = \lambda_0 + 4J/\hbar\Delta$, as shown in Fig.~\ref{fig:F4}. \\

\section*{Appendix I: Creation of highly-excited states}
Bloch oscillations along the synthetic dimension offer a way to controllably populate highly-excited atomic trap states. Indeed, by stopping the shaking at a certain given time it is possible to `freeze' the dynamics. If this is done when the Bloch oscillation is exploring high $\lambda$ states, the resulting state will have a considerable admixture of highly-excited harmonic trap states. In Fig.~\ref{fig:density} we show an example of states obtained after a 5ms TOF expansion with such a technique with $\Delta/2\pi\simeq17$ Hz, $V_0\simeq1.2$ nK and stopping the Bloch evolution when the oscillation has reached the peak (40 ms) and half way (20 ms). By averaging over several different phases of the shaking potential, we can reconstruct the whole set of states that can be obtained for a given Bloch evolution time, this is shown as shaded areas in Fig.~\ref{fig:density} (a) and the column density profiles in Fig. \ref{fig:density} (b) and (c). We observe that the resulting density profiles acquire the characteristic double-lobe structure typical of highly-excited harmonic trap states. Additionally, as the Bloch dynamics evolve, we observe that the distance between the two lobes increases, as expected when populating increasingly higher states. Notably, by controlling the phase of the shaking potential, it is possible to accurately control the shape and position of the final state. This is shown in Fig. \ref{fig:density} (a), where the two solid lines correspond to averages over several runs with the same phase both for the 20 and 40 ms cases. 

We have additionally measured the lifetime of the states created by measuring the number of atoms as a function of time after the Bloch evolution is stopped. We then performed an exponential decay fit, whose decay time sets the lifetime of the state.  Concerning the states shown in Fig. \ref{fig:density} in particular, we have measured that the lifetime for the states produced after 20 ms of shaking potential is $\simeq$1 s, while for those produced after 40 ms is $\simeq$600 ms. Therefore the lifetime of these highly-excited states is sufficiently long to allow one to practically use them. As an example, they would provide a good overlap with a double-well potential enabling new possibilities for trapped atom interferometry \cite{Guarrera2015}.

The fidelity of the excited states could be further improved by decreasing the proportion of the cloud that remains in the low-lying $\lambda$ states, and by decreasing the width with respect to $\lambda$ of the portion that does oscillate. One approach to achieving this is by a combination of using a stronger trap (i.e. increasing $\omega_x$) and/or a lower temperature, both of which reduce the width of the initial Maxwell-Boltzmann distribution. Fig.~\ref{fig:state_fidelity}(a) and (b) show the synthetic space density profile, with $y$-dependence integrated out, for a numerical simulation of a thermal cloud with (a) the typical parameters used elsewhere in our work, and (b) a larger trapping frequency and lower temperature, over a single Bloch oscillation period. As can be seen, in panel (b) the part of the cloud that oscillates is narrower with respect to $\lambda$ than with our more typical parameters in panel (a). Note that, for a small enough temperature, a significant initial condensate fraction may also affect the dynamics. However, we have verified numerically that an initial state with the whole cloud in the $\lambda = 0$ state (i.e. a non-interacting BEC) still undergoes Bloch oscillations under the driving potential.

We also note that the width of the oscillating part of the cloud also depends upon the detuning. In Fig.~\ref{fig:density_lambda_1d}, for the low-power parameters used in the Main Text and a detuning of $\Delta / 2\pi = 2$Hz (panel (a)) and $\Delta / 2\pi = 5$Hz (panel (b)), the density with respect to $\lambda$ is shown close to the peak of the oscillation. The oscillating part of the cloud can be seen to have a full-width at half-maximum (FWHM) of around 8 states in (a). This width depends upon the detuning, with larger detunings yielding smaller widths. For example, in panel (b), the FWHM is around 5 states. Note that these methods to reduce the width of the cloud with respect to $\lambda$ could be extended towards single-site resolution.

Another strategy for improving the fidelity is to optimize the driving potential. As an example, Fig.~\ref{fig:state_fidelity}(c) shows the effect of a different driving potential, $V(x, t) = \kappa x\cos(\omega_Dt)$, upon a lower-temperature thermal cloud ($T = 10$ nK), as calculated numerically using the technique described in Appendix C. Note that this driving potential is the same as that of~\cite{Price2017} (see Eq. 2 therein). As can be seen, the entire cloud undergoes the Bloch oscillation, with no split component. This difference is likely because for the experimental driving potential (Eq. 1 in the Main Text), there are oscillations in the Floquet Hamiltonian matrix elements (e.g. Fig.~\ref{fig:me_low_power}) at low $\lambda$, which include, for example, an effective potential minimum at $\lambda = 0$ which may trap some atoms, leading to the cloud splitting. For the linear driving potential, on the other hand, the Floquet Hamiltonian does not exhibit such oscillations, although there is then a strong $\lambda$-dependence in the hopping elements~\cite{Price2017}. Going further, quantum control approaches could be used to further numerically optimize the driving potential, in order to improve the fidelity of desired state preparation.
\begin{figure}[!t]
	\includegraphics[width=0.48\textwidth]{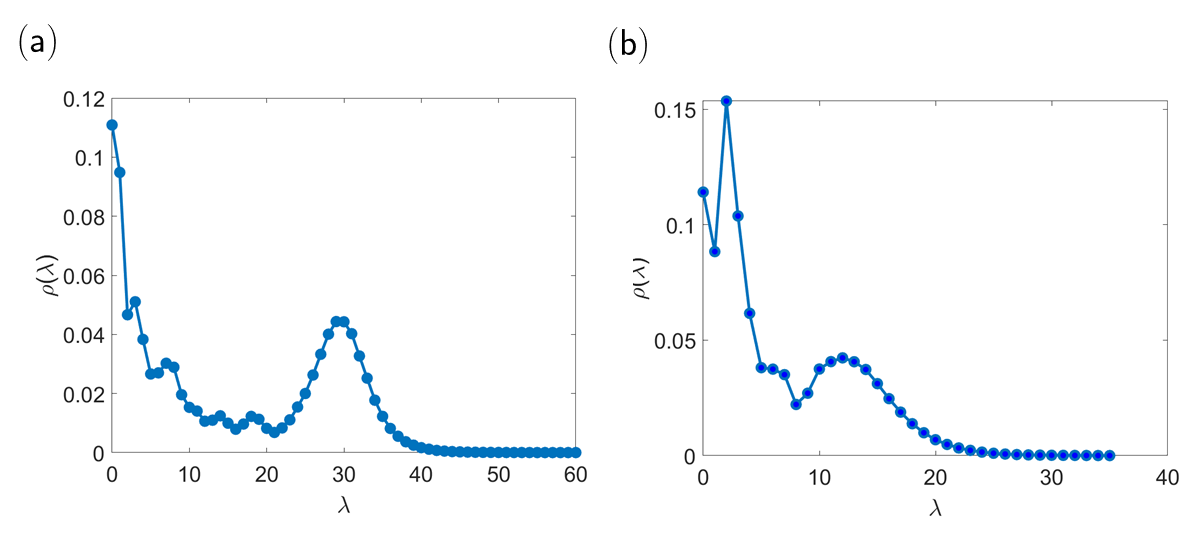}
	\caption{Cloud density with respect to $\lambda$ for the low power parameters used in the Main Text and a detuning of $\Delta / 2\pi = 2$Hz (panel (a)) and $\Delta / 2\pi = 5$Hz (panel (b)). The density is extracted from close to the peak of the oscillation. We clearly see the oscillating part of the cloud, which has a full-width at half-maximum of around 8 states in (a) and around 5 states in (b).}
	\label{fig:density_lambda_1d}
\end{figure}
Finally, another possible strategy for improving the fidelity of excited states would be to employ a two-step protocol. Firstly, the cloud could be placed into an excited trap state without using the driving potential~\cite{Muller2007}. Secondly, the driving potential could be activated, causing Bloch oscillations in the synthetic dimension bulk.

{\it Note Added}: We note that Bloch oscillations of driven dissipative solitons were recently detected in a photonics synthetic dimension~\cite{Englebert2021}. \\

\section*{Acknowledgments}
We would like to acknowledge helpful discussions with Bruno Peaudecerf, David Gu\'{e}ry-Odelin, Tilman Esslinger, Laura Corman and Jean-Philippe Brantut.
We acknowledge financial support by the Royal Society via grants UF160112, RGF\textbackslash{}EA\textbackslash{}180121 and RGF\textbackslash{}R1\textbackslash{}180071, and the EPSRC via grant EP/R021236/1. N.G. is supported by the ERC Starting Grant TopoCold and the Fonds De La Recherche Scientifique (FRS-FNRS, Belgium). G.S. was supported by the ERC Starting Grant TopoCold, the Academy of Finland under project numbers 327293, and the European Union's Horizon 2020 research and innovation programme under the Marie Sk\l{}odowska-Curie grant agreement No 101025211 (TEBLA). \\

A.S. and T.E. carried out the experiment under the guidance and supervision of V.G. and G.B. The theoretical analysis was performed by C.O., under the guidance and supervision of G.S., N.G. and H.M.P. The project was conceived and designed by N.G., H.M.P and G.B. The manuscript was drafted by C.O. and then revised by all authors.

\end{document}